\documentclass[prd,preprintnumbers,superscriptaddress,nofootinbib,longbibliography]{revtex4-1}
\pdfoutput=1
\usepackage{amsmath}
\usepackage{amssymb}
\usepackage{graphicx}
\usepackage{xspace}
\usepackage{units}
\usepackage{slashed}
\usepackage[hyperfootnotes=false]{hyperref}
\usepackage{gensymb}
\usepackage{cancel}
\usepackage{float}
\usepackage[normalem]{ulem}
\usepackage{tabularx}
\usepackage{multirow,array} 

\newcommand{\gev}{\text{GeV}}
\allowdisplaybreaks
\begin{document}

\title{Probing dark sector fermions in Higgs precision studies and direct searches}

\author{Ayres Freitas}
 \email{afreitas@pitt.edu}
\author{Qian Song}
 \email{qis26@pitt.edu}
\affiliation{%
Pittsburgh Particle-physics Astro-physics \& Cosmology Center (PITT-PACC)\\
Department of Physics \& Astronomy, University of Pittsburgh, Pittsburgh, PA 15260, USA
}%

\begin{abstract}
In this paper, we investigate the discovery prospect of simplified fermionic dark sectors models through Higgs precision measurements at $e^+e^-$ colliders and direct searches at hadron colliders. These models extend the Standard Model with two Majorana or Dirac fermions that are singlets, doublets or triplets under the weak SU(2) group. For all models, we consider two scenarios where the lightest new fermion is either stable, or where it decays into other visible final states. For the Higgs precision observables we primarily focus on $\sigma(e^+e^-\to ZH)$, which can deviate from the Standard Model through one-loop corrections involving the new fermions. Deviations of 0.5\% or more, which could be observable at future $e^+e^-$ colliders, are found for TeV-scale dark sector masses. By combining the constraints from the oblique parameters, $\text{Br}(H\to\gamma\gamma)$, and direct production of the new fermions at the LHC, a comprehensive understanding of the discovery potential of these models can be achieved. In both scenarios, there exist some parameter regions where the Higgs precision measurements can provide complementary information to direct LHC searches. 
\end{abstract} 

\maketitle

\section{Introduction}

The discovery of the Higgs boson \cite{ATLAS:2012yve,CMS:2012qbp} opens a new avenue to probe the consistency of the Standard Model (SM) and search for new phenomena. New physics beyond the SM may modify the effective couplings of the Higgs boson relative to the SM expectation. In many well-motivated models, new particles with ${\cal O}$(TeV) masses induce \%-level modifications of the Higgs couplings \cite{Englert:2014uua,Dawson:2013bba,Gu:2017ckc,Peskin:2022pfv}. Deviations of this magnitude can be observed with a future high-luminosity $e^+e^-$ collider operating at a center-of-mass energy of $\sqrt{s} \sim 250$~GeV. There are several proposals for such a machine, including the International Linear Collider (ILC) \cite{ILC:2013jhg,Bambade:2019fyw}, the Future Circular Collider (FCC-ee) \cite{FCC:2018evy}, and the Circular Electron-Positron Collider (CEPC) \cite{CEPCStudyGroup:2018ghi}.

One the other hand, TeV-scale particles can also be searched for directly at the LHC and its high-luminosity extension (HL-LHC). This complementarity between direct searches at high-energy hadron colliders and Higgs precision tests at $e^+e^-$ colliders is highly model-dependent, \emph{i.e.} it must be studied for specific classes of models or simplified models\footnote{Here ``simplified models'' refer to scenarios where the SM is extended by a set of particles that couple to the Higgs boson, but additional particles (that do not couple directly to the Higgs boson) may be needed to construct a UV-complete model.}. In the literature, such comparative investigations have been carried out for a range of models, including the singlet scalar model \cite{ESPPG:2019qin}, Two-Higgs-Doublet models \cite{Chen:2018shg,Chen:2019pkq}, composite Higgs models \cite{Thamm:2015zwa},  supersymmetric extensions of the SM \cite{Bahl:2020kwe}, and fermionic dark matter models \cite{Xiang:2017yfs,Wang:2017sxx}.

Here we extend this body of work by studying the discovery prospects for a class of minimal fermionic dark sector models. The models are constructed by extending the SM by two Majorana or Dirac fermion fields that are singlets, doublet or triplets under weak SU(2). The introduction of more than one fermion field is necessary in order to have renormalizable couplings to the Higgs boson (see \emph{e.g.} Ref.~\cite{Freitas:2015hsa}). For models with one Majorana field, the $e^+e^-$--LHC interplay has also been explored in Ref.~\cite{Xiang:2017yfs}, but in contrast to that work we do not impose dark matter relic density and direct detection constraints, since the dark matter physics may be modified by additional beyond-the-SM (BSM) particles that do not couple to the Higgs boson. As a result, we explore a larger region of parameter space, as well as a larger set of possible (HL-)LHC signatures than Ref.~\cite{Xiang:2017yfs}, which only considered mono-jet constraints.

We assume that the new fermions are charged under a discrete $\mathbb{Z}_2$ symmetry. This prevents mixing between the new fermions and SM fermions, and it implies that the lightest $\mathbb{Z}_2$ fermion is stable and, if it is neutral, a potential dark matter candidate. We will refer to this situation as ``minimal scenarios.'' However, if the minimal dark sector models are embedded in a larger BSM sector, the lightest dark sector fermions could also decay, either into other $\mathbb{Z}_2$-odd BSM particles or into SM particles through a small $\mathbb{Z}_2$-breaking term. We will consider some examples of such ``non-minimal scenarios,'' where we assume that the interaction responsible for the decay of the lightest dark sector fermion is sufficiently small to not affect the Higgs-boson phenomenology, yet large enough to lead to prompt decays at the LHC.

For the dark sector models studied in this article, the most sensitive Higgs precision observable is typically the production cross-section for $e^+e^- \to ZH$, which probes the effective $ZZH$ coupling. This cross-section (henceforth denoted $\sigma_{\rm ZH}$) is modified by the dark sector fermions at the one-loop level. It could be measured with a precision of 1.2\% at ILC, 0.4\% at FCC-ee, and 0.5\% at CEPC. At this level of precision, it is mandatory to take into account radiative corrections in the SM prediction for $\sigma_{\rm ZH}$ \cite{Fleischer:1982af,Kniehl:1991hk,Denner:1992bc,Gong:2016jys,Sun:2016bel,Freitas:2022hyp,Freitas:2023iyx}. We will assume that the SM theory uncertainty for the calculation of $\sigma_{\rm ZH}$ is negligible compared to the experimental measurement precision at ILC/FCC-ee/CEPC. For some simplified models with fermion triplets, modifications of the $H\to\gamma\gamma$ decay rate also become important, and we take those constraints into account.

After briefly reviewing the structure of the minimal fermionic dark sector models in section~\ref{sec:models}, we describe some technical aspects for the calculation of the deviation of $\sigma_{\rm ZH}$ and other phenomenological constraints in section~\ref{sec:calc}. In section.~\ref{sec:result} we show results for the different models in the ``minimal scenario'' where the lightest dark sector fermion is table. In this case, direct production of the new fermions at the LHC lead to signatures with missing energy, and thus they are identical to the expected signals in the Minimal Supersymmetic Standard Model (MSSM). Therefore, we can use supersymmetry (SUSY) searches at the LHC to derive existing and projected bounds on the minimal dark sector models. In section.~\ref{sec:result2}, on the other hand, some examples of ``non-minimal scenarios'' will be discussed. While the predictions for the the $\sigma_{\rm ZH}$ are unchanged, the LHC phenomenology is modified due to decay of the lightest dark sector fermion. We consider two possibilities for this decay, one into SM leptons, which is similar to a SUSY scenario with R-parity violation, and with final states involving photons, which is similar to gauge-mediated SUSY models. There are of course many other possibilities for the lightest dark sector fermion decay, but these two examples serve as representative benchmarks for the expected reach of the (HL-)LHC for such classes of models. Finally, we present our conclusions in section~\ref{sec:concl}.

\section{Models} 
\label{sec:models}
In this section, we introduce two UV complete models that extend the SM with two massive fermionic SU(2) multiplets: one with a singlet and a doublet, and the other with a doublet and a triplet. Both models include a Yukawa interaction term and can be further categorized into five distinct types based on the hypercharge, as discussed in the following sections.

\paragraph{Dirac singlet-doublet model (DSDM) :} This model extends the particle content of the SM by a gauge singlet Dirac field, $\chi_S$, and a SU(2) doublet Dirac field with hypercharge +1/2, $\chi_D = (\chi^+_D,\chi^0_D)$. The mass and Yukawa terms of these fields are given by
\begin{align}
{\cal L}_{\rm DSDM} \supset -m_D\overline{\chi}_D\chi_D-m_S\overline{\chi}_S\chi_S-\bigl(y\,\overline{\chi}_D\chi_S H + \text{h.c}\bigr).
\end{align}
Note that the Yukawa coupling $y$ can be always made real through a complex-phase rotation of $\chi_D$.

The physical spectrum contains a charged Dirac particle $\chi^\pm_D$ with mass $m_D$, and two neutral Dirac mass eigenstates $\chi^0_{l,h}$ with masses $m^0_{l,h}$, where
\begin{align}
    m^0_{l,h} &= \tfrac{1}{2}(m_D+m_S\mp \Delta m_{DS2}), \\ (\Delta m_{XYZ})^2 &\equiv (m_X-m_Y)^2+Z(yv)^2,
\end{align}
and $v \approx 246$~GeV is the vacuum expectation value of the Higgs field, $\langle H \rangle = (0,v/\sqrt{2})$. The mass ordering is $m_l^0 < m_D < m_h^0$.

In the mass eigenstate basis, the interactions of the new fermions with SM bosons read
\begin{align}
    \widehat{\cal L}_{\rm DSDM} \supset &\;\frac{g}{\sqrt{2}} \Bigl[\bigl( \cos\theta\, \overline{\chi^0_l} \gamma^\mu \chi^+_D + \sin\theta\, \overline{\chi^0_h} \gamma^\mu \chi^+_D\bigr)W^-_\mu + \text{h.c.} \Bigr] \notag \\
    &-\frac{g}{2c_W} \Bigl[\cos^2\theta\,\overline{\chi_l^0}\gamma^\mu\chi^0_l + \sin^2\theta\,\overline{\chi_h^0} \gamma^\mu\chi^0_h + \tfrac{1}{2}\sin(2\theta) \bigl(\overline{\chi_l^0}\gamma^\mu\chi^0_h + \overline{\chi_h^0}\gamma^\mu\chi^0_l\bigr) + (2s_W^2-1)\,\overline{\chi^+_D}\gamma^\mu\chi^+_D\Bigr]Z_\mu \notag \\
    &-e\,\overline{\chi^+_D}\gamma^\mu\chi^+_DA_\mu -\frac{y}{\sqrt{2}}\Bigl[ \sin(2\theta) \bigl(\overline{\chi^0_h}\chi^0_h - \overline{\chi^0_l}\chi^0_l\bigr) + \cos(2\theta) \bigl(\overline{\chi^0_l}\chi^0_h + \overline{\chi^0_h}\chi^0_l\bigr)\Bigr]h\,,
\end{align}
where $s_W$ and $c_W$ are the sine and cosine of the Weinberg angle, respectively, and
\begin{align}
    \sin\theta = s_{DS2}, \qquad s^2_{XYZ} \equiv \frac{1}{2}\biggl(1+\frac{m_X-m_Y}{\Delta m_{XYZ}}\biggr).
    \label{eq:mixang}
\end{align}
If there is a $\mathbb{Z}_2$ symmetry, $\chi^0_l$ is stable and leads to missing-energy signatures at colliders, whereas $\chi^\pm_D$ and $\chi^0_h$ decay into $\chi^0_l$ via emssions of on- or off-shell SM bosons,
\begin{align}
    \chi^\pm_D \to W^{(*)\pm}\chi^0_l, \quad
    \chi^0_h \to Z^{(*)}\chi^0_l, \,h^{(*)}\chi^0_l.
\end{align}
For more details and some phenomenological studies of this model, see Refs.~\cite{Freitas:2015hsa,Fedderke:2015txa,Yaguna:2015mva}.

\bigskip
\paragraph{Majorana singlet-doublet model (MSDM):} In contrast to the DSDM, this model has a Majorana gauge singlet field, which we also denote as $\chi_S$. In the following, Majorana fields are denoted in terms of 4-component Majorana spinors. The mass and Yukawa sector of this model reads
\begin{align}
{\cal L}_{\rm MSDM} \supset -m_D\overline{\chi}_D\chi_D-\tfrac{1}{2}m_S\overline{\chi}_S\chi_S-\bigl(y\,\overline{\chi}_D\chi_S H + \text{h.c}\bigr).
\end{align}
In addition to the charged Dirac particle $\chi^\pm_D$, this model has three neutral Majorana mass eigenstates, $\chi^0_{l,m,h}$, with masses
\begin{align}
m^0_{l,h} &= \tfrac{1}{2}(m_D+m_S\mp \Delta m_{DS4}),
\qquad m^0_m = m_D,
\end{align}
and, for sufficiently large values of $m_D$, the ordering $m^0_l < m^0_m < m^0_h$.
Their interactions are given by
\begin{align}
    \widehat{\cal L}_{\rm MSDM} \supset &\;\frac{g}{2} \Bigl[\bigl( -\cos\theta\, \overline{\chi^0_l} \gamma^\mu \chi^+_D + \sin\theta\, \overline{\chi^0_h} \gamma^\mu \chi^+_D + i\,\overline{\chi^0_m} \gamma^\mu \chi^+_D \bigr)W^-_\mu + \text{h.c.} \Bigr] \notag \\
    &+\frac{g}{2c_W} \Bigl[-i\,\cos\theta\,\overline{\chi_l^0}\gamma^\mu\chi^0_m + i\,\sin\theta\,\overline{\chi_h^0}\gamma^\mu\chi^0_m + (1-2s_W^2)\,\overline{\chi^+_D}\gamma^\mu\chi^+_D\Bigr]Z_\mu \notag \\
    &-e\,\overline{\chi^+_D}\gamma^\mu\chi^+_DA_\mu +\frac{y}{2}\Bigl[ \sin(2\theta) \bigl(\overline{\chi^0_l}\chi^0_l - \overline{\chi^0_h}\chi^0_h\bigr) + \cos(2\theta) \bigl(\overline{\chi^0_l}\chi^0_h + \overline{\chi^0_h}\chi^0_l\bigr)\Bigr]h\,,
\end{align}
where $\sin\theta = s_{DS4}$, with the symbol $s_{XYZ}$ defined in eq.~\eqref{eq:mixang}.

The MSDM has been studied extensively in the literature \cite{Carena:2004ha,Mahbubani:2005pt,DEramo:2007anh,Enberg:2007rp,Cohen:2011ec,Cheung:2013dua,Calibbi:2015nha,Freitas:2015hsa,Banerjee:2016hsk,Abe:2017glm,Xiang:2017yfs,Bhattacharya:2018fus}. Ref.~\cite{Voigt:2017vfz} explored some of the interplay between the LHC and $e^+e^-$ colliders in probing this model, with a focus on Higgs observables at both colliders.

\bigskip
\paragraph{Dirac doublet-triplet model:} Extensions of the SM with one SU(2) doublet and one SU(3) triplet Dirac field allow some freedom in the assignment of hypercharges. Here we consider two options, one where the triplet field has hypercharge $-1$, and the other where the triplet field has hypercharge 0. There are very few existing phenomenological studies of these models in the literature \cite{Freitas:2015hsa}.

\medskip
\paragraph*{Dirac doublet-triplet model with triplet hypercharge $-1$ (DDTM1):} To form a Yukawa coupling, the doublet field must have hypercharge $-1/2$ in this case. The field components can be written as
\begin{align}
 \chi_D = \begin{pmatrix}\chi_D^0\\ \chi_D^- \end{pmatrix},\qquad \chi_T = \begin{pmatrix}
\chi_T^-/\sqrt{2} & \chi_T^0 \\ \chi_T^{--} & -\chi_T^-/\sqrt{2}\end{pmatrix}.\label{eq:dtm1}
\end{align}
The mass terms and Yukawa interactions are
\begin{align}
{\cal L}_{\rm DDTM1} \supset -m_D\overline{\chi}_D\chi_D-m_T\,\text{Tr}[\overline{\chi}_T\chi_T]-\bigl(y\,\overline{\chi}_D\chi_T H + \text{h.c}\bigr). \label{eq:LagDDTM1}
\end{align}
The physical spectrum contains two neutral Dirac particles $\chi^0_{l,h}$, two singly charged Dirac particles $\chi^\pm_{l,h}$, and one doubly charged field $\chi^{\pm\pm}_T$. Their masses are given by
\begin{align}
    m^0_{l,h} = \tfrac{1}{2}(m_D+m_T\mp \Delta m_{DT2}), \qquad
    m^\pm_{l,h} = \tfrac{1}{2}(m_D+m_T\mp \Delta m_{DT1}), \qquad
\qquad m^{\pm\pm} = m_T,
\end{align}
and their interactions read
\begin{align}
    \widehat{\cal L}_{\rm DDTM1} \supset &\;g \Bigl\{\Bigl[ (\tfrac{1}{\sqrt{2}}cc'-ss')\, \overline{\chi^0_l} \gamma^\mu \chi^-_l + (\tfrac{1}{\sqrt{2}}ss'-cc')\, \overline{\chi^0_h} \gamma^\mu \chi^-_h + (\tfrac{1}{\sqrt{2}}sc'+cs')\, \overline{\chi^0_h} \gamma^\mu \chi^-_l \notag \\
    &\qquad + (\tfrac{1}{\sqrt{2}}cs'+sc')\, \overline{\chi^0_l} \gamma^\mu \chi^-_h -s'\, \overline{\chi^-_l}\gamma^\mu\chi^{--}_T + c'\, \overline{\chi^-_h}\gamma^\mu\chi^{--}_T \Bigr]W^+_\mu + \text{h.c.} \Bigr\} \notag \\
    &-\frac{g}{2c_W} \Bigl[(c^2-2)\,\overline{\chi_l^0}\gamma^\mu\chi^0_l + (s^2-2)\,\overline{\chi_h^0}\gamma^\mu\chi^0_h + sc \bigl(\overline{\chi_l^0}\gamma^\mu\chi^0_h + \overline{\chi_h^0}\gamma^\mu\chi^0_l\bigr) \notag \\
    &\qquad + (c'^2-2s_W^2)\,\overline{\chi^-_l}\gamma^\mu\chi^-_l + (s'^2-2s_W^2)\,\overline{\chi^-_h}\gamma^\mu\chi^-_h + s'c' \bigl(\overline{\chi_l^-}\gamma^\mu\chi^-_h + \overline{\chi_h^-}\gamma^\mu\chi^-_l\bigr) \Bigr] \notag \\
    &\qquad + 2(1-2s_W^2)\, \overline{\chi^{--}_T}\gamma^\mu \chi^{--}_T \Bigr] Z_\mu \notag \displaybreak[0] \\
    &+ e\Bigl[ \overline{\chi^-_l}\gamma^\mu\chi^-_l + \overline{\chi^-_h}\gamma^\mu\chi^-_h + 2\,\overline{\chi^{--}_T}\gamma^\mu \chi^{--}_T \Bigr] A_\mu
    \notag \displaybreak[0] \\
    &-\frac{y}{\sqrt{2}}\Bigl[ \sin(2\theta) \bigl(\overline{\chi^0_h}\chi^0_h - \overline{\chi^0_l}\chi^0_l\bigr) + \cos(2\theta) \bigl(\overline{\chi^0_l}\chi^0_h + \overline{\chi^0_h}\chi^0_l\bigr)\Bigr]h \notag \\
    &+\frac{y}{2}\Bigl[ \sin(2\theta') \bigl(\overline{\chi^-_h}\chi^-_h - \overline{\chi^-_l}\chi^-_l\bigr) + \cos(2\theta') \bigl(\overline{\chi^-_l}\chi^-_h + \overline{\chi^-_h}\chi^-_l\bigr)\Bigr]h\,,
\end{align}
where $s\equiv \sin\theta = s_{DT2}$, $c\equiv \cos\theta$, $s'\equiv \sin\theta' = -s_{DT1}$, $c'\equiv \cos\theta'$.

\medskip
\paragraph*{Dirac doublet-triplet model with triplet hypercharge 0 (DDTM0):} This model shares some similarities with the DDTM1 model, but now the doublet field has hypercharge $+1/2$, and the field components are given by
\begin{align} \chi_D = \begin{pmatrix}\chi_D^+\\ \chi_D^0 \end{pmatrix},\qquad \chi_T = \begin{pmatrix}
\chi_T^0/\sqrt{2} & \chi_T^+ \\ \chi_T'^{-} & -\chi_T^0/\sqrt{2}\end{pmatrix}.\label{eq:dtm0}
\end{align}
The mass and Yukawa terms have the same form as in \eqref{eq:LagDDTM1}.

The mass term and Yukawa interactions are same as Eq.\ref{eq:LagDDTM1}. The model contains two neutral Dirac particles $\chi^0_{l,h}$ and three singly charged Dirac particles $\chi^\pm_{l,m,h}$, with the following masses:
\begin{align}
    \tilde{m}^0_{l,h} = \tfrac{1}{2}(m_D+m_T\mp \Delta m_{DT1}), \qquad
    \tilde{m}^\pm_{l,h} = \tfrac{1}{2}(m_D+m_T\mp \Delta m_{DT2}), \qquad
\qquad m^{\pm}_m = m_T,
\end{align}
Since we assume a $\mathbb{Z}_2$ symmetry in the dark sector, the lightest fermion should be neutral, \emph{i.e.} we require $|m_l^0|<|m_l^\pm|$. For this model this can only be achieved if $m_D$ and $m_T$ have opposite signs \cite{Freitas:2015hsa}. For the phenomenological studies in section~\ref{sec:DirDT0Model} we show results in terms of ``sign-less'' masses, $m_{l,h}^{0,\pm} \equiv |\tilde{m}_{l,h}^{0,\pm}|$.

The interactions of the new fermions read
\begin{align}
    \widehat{\cal L}_{\rm DDTM0} \supset &\;g \Bigl\{\Bigl[ (\tfrac{1}{\sqrt{2}}cc'-ss')\, \overline{\chi^0_l} \gamma^\mu \chi^+_l + (\tfrac{1}{\sqrt{2}}ss'-cc')\, \overline{\chi^0_h} \gamma^\mu \chi^+_h + (\tfrac{1}{\sqrt{2}}sc'+cs')\, \overline{\chi^0_h} \gamma^\mu \chi^+_l \notag \\
    &\qquad + (\tfrac{1}{\sqrt{2}}cs'+sc')\, \overline{\chi^0_l} \gamma^\mu \chi^+_h -s\, \overline{\mbox{$\chi^-_m$}}\gamma^\mu\chi^0_l + c\, \overline{\mbox{$\chi^-_m$}}\gamma^\mu\chi^0_h \Bigr]W^-_\mu + \text{h.c.} \Bigr\} \notag \displaybreak[0] \\
    &-\frac{g}{2c_W} \Bigl[c^2\,\overline{\chi_l^0}\gamma^\mu\chi^0_l + s^2\,\overline{\chi_h^0}\gamma^\mu\chi^0_h + sc \bigl(\overline{\chi_l^0}\gamma^\mu\chi^0_h + \overline{\chi_h^0}\gamma^\mu\chi^0_l\bigr) \notag \\
    &\qquad + (c'^2-2c_W^2)\,\overline{\chi^+_l}\gamma^\mu\chi^+_l + (s'^2-2c_W^2)\,\overline{\chi^+_h}\gamma^\mu\chi^+_h + s'c' \bigl(\overline{\chi_l^+}\gamma^\mu\chi^+_h + \overline{\chi_h^+}\gamma^\mu\chi^+_l\bigr) \Bigr] \notag \\
    &\qquad + 2c_W^2\, \overline{\mbox{$\chi^-_m$}} \gamma^\mu \chi^{-}_m \Bigr] Z_\mu \notag \displaybreak[0] \\
    &+ e\Bigl[ -\overline{\chi^+_l}\gamma^\mu\chi^+_l - \overline{\chi^+_h}\gamma^\mu\chi^+_h + \overline{\mbox{$\chi^-_m$}}\gamma^\mu \chi^-_m \Bigr] A_\mu
    \notag \displaybreak[0] \\
    &+\frac{y}{2}\Bigl[ \sin(2\theta) \bigl(\overline{\chi^0_h}\chi^0_h - \overline{\chi^0_l}\chi^0_l\bigr) + \cos(2\theta) \bigl(\overline{\chi^0_l}\chi^0_h + \overline{\chi^0_h}\chi^0_l\bigr)\Bigr]h \notag \\
    &-\frac{y}{\sqrt{2}}\Bigl[ \sin(2\theta') \bigl(\overline{\chi^+_h}\chi^+_h - \overline{\chi^+_l}\chi^+_l\bigr) + \cos(2\theta') \bigl(\overline{\chi^+_l}\chi^+_h + \overline{\chi^+_h}\chi^+_l\bigr)\Bigr]h\,,
\end{align}
where $s\equiv \sin\theta = s_{DT1}$,  and $s'\equiv \sin\theta' = -s_{DT2}$, \emph{i.e.} the values of $\theta$ and $\theta'$ are reversed compared to the DDTM1 model.

\bigskip
\paragraph{Majorana doublet-triplet model (MDTM):} If the triplet has hypercharge 0, it can also be a Majorana field (similar to the wino in supersymmetry). Again using 4-component Majorana spinor fields, the Largrangian for the mass and Yukawa terms reads
\begin{align}
{\cal L}_{\rm MDTM} \supset -m_D\overline{\chi}_D\chi_D-\tfrac{1}{2}m_T\,\text{Tr}[\overline{\chi}_T\chi_T]-\bigl(y\,\overline{\chi}_D\chi_T H + \text{h.c}\bigr). \label{eq:LagMDTM}
\end{align}
The physical mass spectrum of this model encompasses two charged Dirac states $\chi^\pm_{l,h}$ and three neutral Majorana states $\chi^0_{l,m,h}$. At tree-level, their masses are
\begin{align}
    m^0_{l,h} = m^\pm_{l,h} = \tfrac{1}{2}(m_D+m_T\mp \Delta m_{DT2}), \qquad
\qquad m^0_m = m_D.
\end{align}
Unless $m_D$ is very small, one has the mass ordering $m^0_l < m^0_m < m^0_h$. The degeneracy between $m_l^0$ and $m_l^\pm$ is lifted through one-loop corrections \cite{Giudice:1995np,Cheng:1998hc,Feng:1999fu,Gherghetta:1999sw}, producing a small positive contribution to $m_l^\pm-m_l^0$, so that the lightest state is charge-neutral.
The relevant interactions in the mass basis are
\begin{align}
    \widehat{\cal L}_{\rm MDTM} \supset &\;\frac{g}{2} \Bigl\{\Bigl[ (1+\sin^2\theta)\, \overline{\chi^0_l} \gamma^\mu \chi^+_l -(1+\cos^2\theta)\, \overline{\chi^0_h} \gamma^\mu \chi^+_h +
    \tfrac{1}{2}\sin(2\theta) \bigl( \overline{\chi^0_l} \gamma^\mu \chi^+_h - \overline{\chi^0_h} \gamma^\mu \chi^+_l \bigr) \notag \\
    &\qquad + i\,\cos\theta\,\overline{\chi^0_m} \gamma^\mu \chi^+_l + i\,\sin\theta\,\overline{\chi^0_m} \gamma^\mu \chi^+_h \Bigr]W^-_\mu + \text{h.c.} \Bigr\} \notag \\
    % check whether sin/cos in line above are correct
    &+\frac{g}{2c_W} \Bigl[i\,\cos\theta\,\overline{\chi_l^0}\gamma^\mu\chi^0_m - i\,\sin\theta\,\overline{\chi_h^0}\gamma^\mu\chi^0_m \notag \\
    &\qquad - (\cos^2\theta-2c_W^2)\,\overline{\chi^+_l}\gamma^\mu\chi^+_l - (\sin^2\theta-2c_W^2)\,\overline{\chi^+_h}\gamma^\mu\chi^+_h - \sin\theta\cos\theta \bigl(\overline{\chi_l^+}\gamma^\mu\chi^+_h + \overline{\chi_h^+}\gamma^\mu\chi^+_l\bigr) \Bigr]  \Bigr] Z_\mu \notag \displaybreak[0] \\
    &- e\Bigl[ \overline{\chi^+_l}\gamma^\mu\chi^+_l + \overline{\chi^+_h}\gamma^\mu\chi^+_h \Bigr] A_\mu
    \notag \displaybreak[0] \\
    &+\frac{y}{\sqrt{2}}\Bigl[ \sin(2\theta) \bigl(\tfrac{1}{2}\overline{\chi^0_l}\chi^0_l - \tfrac{1}{2}\overline{\chi^0_h}\chi^0_h + \overline{\chi^+_l}\chi^+_l - \overline{\chi^+_h}\chi^+_h\bigr) + \cos(2\theta) \bigl(\tfrac{1}{2}\overline{\chi^0_l}\chi^0_h + \tfrac{1}{2}\overline{\chi^0_h}\chi^0_l - \overline{\chi^+_l}\chi^+_h - \overline{\chi^+_h}\chi^+_l\bigr)\Bigr]h\,,
\end{align}
where $\sin\theta = s_{DT2}$. See Ref.~\cite{Dedes:2014hga,Freitas:2015hsa,Cai:2016sjz,Xiang:2017yfs} for more information and further studies of this model.

\section{Calculation of experimental sensitivity and constraints}
\label{sec:calc}
In this section, we start by introducing the computational techniques for calculating the deviation of $\sigma_{\rm ZH}$ and then discuss all constraints for phenomenological study, which include oblique parameters, Higgs diphoton decay branching fraction, and collider searches at the (HL-)LHC.

\subsection{NLO corrections to $\sigma_{\rm ZH}$}
\begin{figure}[t]
\centering
\includegraphics[width=0.6\textwidth]{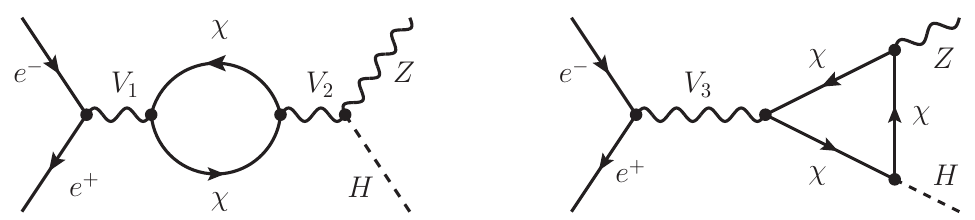}
\vspace{-1em}
\caption{Self-energy (left) and vertex (right) Feynman diagrams with new fermions, denoted as $\chi$. $V_{i}=\gamma,Z$. }
\label{fig:FDMeezh}
\end{figure}

The deviation of $\sigma_{\rm ZH}$ due to new fermions is defined as 
\begin{align}
\delta = \biggl|\frac{\sigma^{\text{FDS}}_{\rm ZH}-\sigma^{\text{SM}}_{\rm ZH}}{\sigma^{\text{SM}}_{\rm ZH}}\biggr|, \label{eq:delta}
\end{align}
where ``FDS'' denotes any of the models introduced in the previous section, with extend the SM by a set of dark sector fermions. Both integrated cross sections are computed assuming unpolarized electron-positron beams. The inclusion of polarized beams does not introduce any changes, as all dark matter fermions are considered to be vector fermions. The Standard Model contribution, $\sigma^{\text{SM}}_{\rm ZH}$, takes into account one-loop EW as well as fermionic two-loop electroweak corrections, and the corresponding result is obtained from Ref.~\cite{Freitas:2022hyp}. On the other hand, the numerator of eq.~\eqref{eq:delta}, which involves contributions from dark sector fermions, is calculated at one-loop order. At the one-loop level, the inclusion of new fermions contributes to $\sigma^{\text{FDM}}_{\rm ZH}$ through self-energy and vertex contributions. The corresponding Feynman diagrams illustrating these contributions are depicted in Fig.\ref{fig:FDMeezh}.

For the generation of the corresponding one-loop amplitudes, we use the package {\tt FeynArts 3.11} \cite{Hahn:2000kx}. All models introduced in the previous section have been implemented in {\tt FeynArts} model files\footnote{The model files are available from the authors upon request, written in two different ways. One utilizes private code, while the other implements {\tt FeynRules} packages\cite{feynrules}, serving as a cross-check. }. For the computation of the cross-section, {\tt FeynCalc 9} \cite{Shtabovenko:2020gxv} and a private Mathematica code have been used, and it has been checked that the two sets of results agree.

In this work, we employ the on-shell renormalization scheme. The dark sector fermions, serving as intermediate states, require no renormalization for their masses, fields, or Yukawa coupling. However, they do influence the self-energy of the SM bosons, leading to extra contributions to the boson masses, fields, mixing angle, and electromagnetic coupling from these new fermion interactions. These corrections to the SM counterterms are also computed using the {\tt FeynArts} model files mentioned above.

\subsection{Oblique parameters}
Assuming that the new fermions are heavier than the W and Z boson, they indirectly contribute to the propagators of gauge bosons through off-shell loop effects, i.e.\ the so-called oblique corrections\cite{Peskin:1990zt,Peskin:1991sw,Maksymyk:1993zm}. The effects of oblique corrections are parameterized by six oblique parameters: $S,T,U,V,X$ and $W$. Among these, the first three parameters are strongly constrained by electroweak precision observables. Furthermore, we fix $U=0$, which is motivated by the fact that $U$ is suppressed by an additional factor $M_{\text{new}}^2/M_Z^2$ compared to $S$ and $T$ where $M_{\text{new}}$ is the energy scale of new physics. This suppression can also be understood from an effective field theory (EFT) perspective: $S$ and $T$ correspond to dimension-6 operators $H^\dagger W_{\mu\nu}^a\sigma^aHB^{\mu\nu}$ and $H^\dagger(D_\mu H)(D^\mu H)^\dagger H$, respectively, while the operator contributing to U in the lowest order is a dimension-8 operator $H^\dagger W_{\mu\nu}^a\sigma^aHH^\dagger W^{b\mu\nu}\sigma^b H$ \cite{Han:2008es}. 

The definitions of $S$ and $T$ at one-loop level are
\begin{align}
\frac{\alpha}{4s_W^2c_W^2} S &
= \frac{\Pi_{ZZ}^{\text{new}}(M_Z^2)-\Pi_{ZZ}^{\text{new}}(0)}{M_Z^2} -\frac{c_W^2-s_W^2}{c_Ws_W} \frac{\Pi_{Z\gamma}^{\text{new}}(M_Z^2)}{M_Z^2} - \frac{\Pi_{\gamma\gamma}^{\text{new}}(M_Z^2)}{M_Z^2} \\
\alpha T
&= -\frac{s_W^2}{c_W^2M_Z^2}\Sigma_{AA}^\text{new}(0)+\frac{1}{M_W^2}\Sigma_{WW}^\text{new}(0)-\frac{2s_W}{c_WM_Z^2}\Sigma_{ZA}^{\rm new}(0)-\frac{1}{M_Z^2}\Sigma_{ZZ}^{\rm new}(0) 
\end{align}
where $\Sigma^{\rm new}$ stands for the transversal gauge-boson self-energy corrections from new physics, and $\Pi(p^2) = \Sigma(p^2)/p^2$. Under the assumption $U=0$, the numerical values of $S$ and $T$ from a multi-parameter fit at 95\% CL are \cite{pdg2022}
\begin{align}
S=-0.01\pm 0.14,~~T=0.04\pm 0.12. \label{eq:STvals}
\end{align}
As explained in the previous subsection, we compute the corrections to $S$ and $T$ from the models in section~\ref{sec:models} using {\tt FeynArts}, {\tt FeynCalc} and a private Mathematica code. For the Majorana models (MSDM, MDTM), the $T$ parameter correction is zero and the $S$ parameter contribution tends to be numerically small and within the electroweak precision limits. On the other hand, the bounds in eq.~\eqref{eq:STvals} significantly impact the allowed parameter space for the Dirac models (DSDM, DDTM1, DDTM0), as will be shown in section~\ref{sec:result}.

\subsection{Higgs decays}\label{sec:higgsdecay}
In the doublet-triplet models, the decay rate of Higgs to di-photons is changed at one-loop level due to the presence of virtual charged fermions. The decay ratio with respect to the SM rate is given by
\begin{align}
R_\gamma = \frac{\Gamma(h\to \gamma\gamma)}{\Gamma_{\text{SM}}(h\to \gamma\gamma)} = \Big|1+\frac{A_\chi}{A_\text{SM}} \Big|^2
\end{align}
The one-loop corrections for the SM, $A_\text{SM}$, and the fermionic dark sector, $A_\chi$, are defined as
\begin{align}
A_\text{SM} = \sum_f N_cA_f^2A_F(\tau_F)+A_B(\tau_W), \qquad A_\chi=\sum_\chi Q_\chi^2 y_\chi \frac{v}{m_\chi}A_F(\tau_\chi)
\end{align}
with $\tau_i=m_H^2/4m_i^2$, where $m_H$ is the Higgs mass. The loop functions $A_{f,B}$ for $\tau\leq 1$ are given by \cite{Shifman:1979eb,Gastmans:2011wh,
Huang:2011yf,Freitas:2015hsa}
\begin{align}
&A_F(\tau) = \frac{2}{\tau^2}\Big\{\tau+(\tau-1)\arcsin^2\sqrt{\tau} \Big\}, \notag \\
&A_B(\tau) = -\frac{1}{\tau^2}\Big\{2\tau^2+3\tau+(6\tau-3)\arcsin^2\sqrt{\tau} \Big\}
\end{align}
For the three doublet-triplet models, the expression for $A_\chi$ is given by
\begin{align}
A_\chi&=\frac{v^2y^2}{2(m_h^\pm-m_l^\pm)}\Big( \frac{1}{m_h^\pm}A_F(\tau_{\chi_h^\pm})-\frac{1}{m_l^\pm}A_F(\tau_{\chi_l^\pm}) \Big) \times \left\{ \begin{array}{rl}
1 & \mbox{for DDTM1} \\
 & \\
2 & \mbox{for DDTM0, MDTM}
\end{array}\right.
\end{align}

% Run-1 meansurement result \cite{CMS:2014afl} \cite{ATLAS:2014cnc}, 

The current measurements of $R_{\gamma}$ from the ATLAS and CMS experiments at 95\% CL are $R_{\gamma}=1.04^{+0.20}_{-0.18}$ \cite{ATLAS:2022tnm} and $R_{\gamma}=1.12\pm 0.18$ \cite{CMS:2021kom}, respectively. The projected result at HL-LHC can be found in Ref.~\cite{Cepeda:2019klc}, in which the uncertainties are expected to be reduced to 8\% by combining the gluon-fusion and $b\bar{b}$ annihilation channels. The expected precision at FCC-ee is $R_{\gamma}=1\pm 0.09$\cite{FCC:2018evy}, which is comparatively lower than the projected HL-LHC result. Therefore we do not show constraints on $R_\gamma$ from FCC-ee or other future $e^+e^-$ in our plots below.

\subsection{Collider searches}
In addition to indirect constraints from oblique parameters and Higgs precision physics, dark sector fermions with masses of $\mathcal{O}$(TeV) or less could be directly produced at the LHC. Due to the $\mathbb{Z}_2$ symmetry, fermionic dark matter must be produced in pairs, and particles heavier than $\chi_l^0$ eventually decay into $\chi_l^0$. The relevant production channels include
\begin{align}
&q\Bar{q}'\to W^{*\pm} \to \chi^\pm (\to\chi_l^0W^{*\pm})+\chi^0(\to \chi_l^0+Z^{*}/H), \notag \\
&q\Bar{q}' \to Z^{*} \to \chi^0(\to \chi_l^0+Z^{*}/H)+\chi^0(\to \chi_l^0+Z^{*}/H) \label{eq:lhcprod} \\
&q\Bar{q}' \to Z^{*} \to \chi^\pm (\to\chi_l^0W^{\pm*})+\chi^\mp(\to\chi_l^0W^{\mp*}),  \notag
\end{align}
where $\chi^{\pm,0}$ denote the heavier charged and neutral particles, which decay to the lightest neutral particle through W, Z or H emission\footnote{The heavier neutral particles can also decay to $\gamma+\chi_l^0$ through a loop-induced interaction, which is suppressed and thus not included.}. The first production channel typically leads to the strongest constraints since the production cross section is largest for the models considered here. The Z, W and Higgs can decay either leptonically or hadronically, which lead to the signatures with hadronic, semi-leptonic and fully leptonic final states plus missing energy if the $\mathbb{Z}_2$ symmetry is exact. 

The experimental signatures for the processes in \eqref{eq:lhcprod} are very similar to electroweak production of charginos and neutralinos in the Minimal Supersymmetric Standard Model (MSSM). Therefore, we translate bounds from existing supersymmetry (SUSY) searches at the LHC, as well as projections for upcoming HL-LHC. The charged fermions in our models share similar properties to charginos, and the Majorana fermions are similar to the neutralinos. As a result, the exclusion limits from SUSY searches can be directly applied to the Majorana models (MSDM, MDTM). Additionally, we can obtain exclusion limits for models involving neutral Dirac fermions through the following recast
\begin{align}
\mathcal{S}^{\rm Dirac} = \mathcal{S}^{\rm MSSM} \times \frac{\sigma^{\rm Dirac}(q\Bar{q}'\to\chi^{0,\pm}\chi^{0,\mp})}{\sigma^{\rm MSSM}(q\Bar{q}'\to\chi^{0,\pm}\chi^{0,\mp})}
\end{align}
where $\mathcal{S}^{\rm Dirac,MSSM}$ denote the signal significance of the Dirac model and MSSM, respectively. The cross sections correspond to the processes in \eqref{eq:lhcprod}. 

The searches for R-parity conserving SUSY at colliders can be categorized according to different final states. Fully hadronic final states benefit from large SM gauge boson decay branching ratios, but they require large amounts of missing energy to distinguish the signal from the SM background. Thus these search channels are sensitive to scenarios with large mass splitting. Multi-lepton final states are sensitive to scenarios with moderate mass splitting, while this class of searches fails in the compressed mass scenario, since the leptons from the decays become too soft to pass the event selection trigger. In such case, an energetic jet from initial state radiation can help to enhance the detectability of the signal. The final state particles recoil against the ISR jet, i.e.\ the missing transverse momentum is of the same order as the jet, which helps to suppress backgrounds. According to Ref.~\cite{Ismail:2016zby}, the final state with a soft photon, jet and missing energy can also improve the sensitivity for compressed mass scenarios. All categories have been taken into account, as we will show in the Sec.\ref{sec:result}. 

In the context of the non-minimal scenarios, the final state signature will be similar to the one in minimal scenario, but involves additional decay products from the unstable lightest fermion. As discussed in the introduction, final states signature with four leptons and di-photons are considered as benchmarks, allowing us to directly implement the exclusion limits obtained from searches for R-parity violating SUSY and gauge-mediated SUSY. The collider searches for RPV SUSY have studied in Refs.\cite{ATLAS:2021yyr,RPVref2,RPVref3}, and the collider search for gauge-mediated SUSY can be found in Refs.\cite{GGMref1,GGMref2,ATLAS:2018nud,GGMref4,GGMref5}. In these searches, the heavy charginos and neutralinos are assumed to be a mixture of winos and Bino. Encountering the other possibilities of heavy charginos and neutralinos, we take advantage of the upper limit on the signal cross section. A more detailed analysis will be presented in Sec.\ref{sec:result2}.

\section{Results for minimal scenarios with stable dark sector fermions} \label{sec:result}

The following input parameters are used for the numerical evaluation:
\begin{alignat}{3}
& m_W^{\rm exp} = 80.379~\text{GeV}\quad &&\Rightarrow\quad &&m_W = 80.352~\text{GeV}, \nonumber \\
&m_Z^{\rm exp} = 91.1876~\text{GeV}  &&\Rightarrow\quad &&m_Z = 91.1535~\text{GeV}, \nonumber \\[1ex]
&m_H = 125.1~\text{GeV},&&  &&  m_t = 172.76~\text{GeV}, \nonumber \\ 
&\alpha^{-1} = 137.036,&&  && \Delta \alpha = 0.059, \nonumber \\
& \sqrt{s} = 240~\text{GeV}. \label{eq:input} 
\end{alignat}
where $\sqrt{s}$ represents the center-of-mass energy, and the masses of all other fermions are set to be 0.

\subsection{Dirac singlet-doublet model}
\label{sec:DirSDModel}
\begin{figure}[t]
    \centering
    \includegraphics[width=0.65\textwidth]{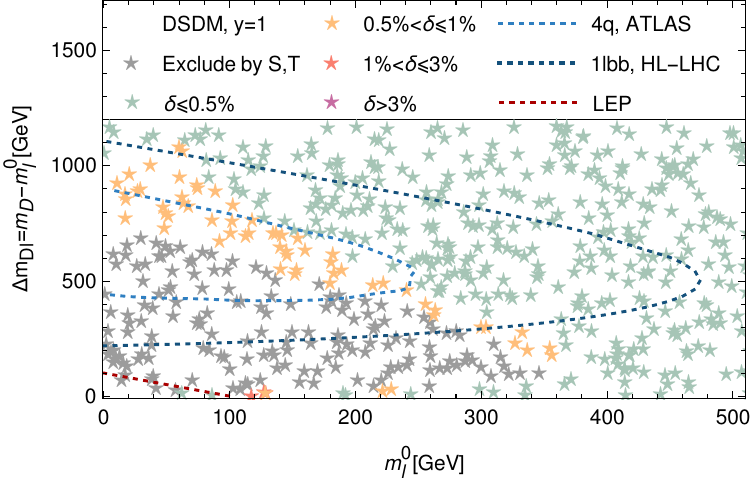}
    \\[1em]
    \includegraphics[width=0.65\textwidth]{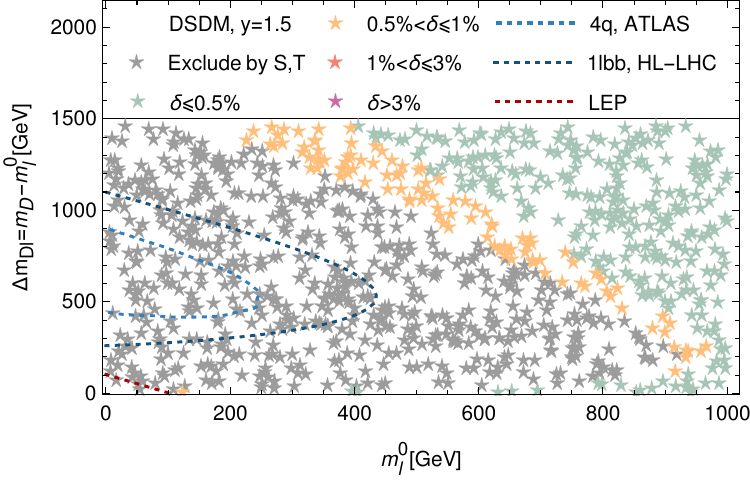}
    \caption{Parameter scan result for DSDM with $y=1$ (upper) and $y=1.5$ (lower). The dashed lines are the 95\% CL exclusion contour based on Refs.~\cite{ATLAS:2021yqv} (``4q, ATLAS''), \cite{ATLAS:2018diz}(``1lbb, HL-LHC'') and \cite{DELPHI:2003uqw} (``LEP''), respectively. }
    \label{fig:DirSDscan1}
\end{figure}

\begin{figure}[t]
    \centering
    \includegraphics[width=0.65\textwidth]{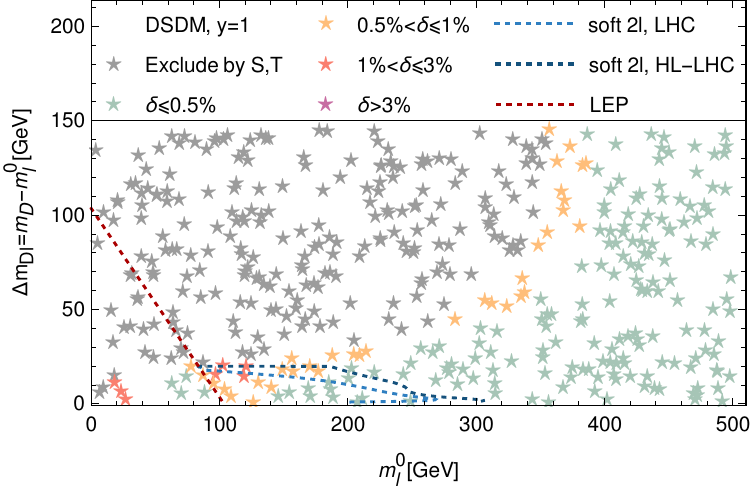}
    \\[1em]
    \includegraphics[width=0.65\textwidth]{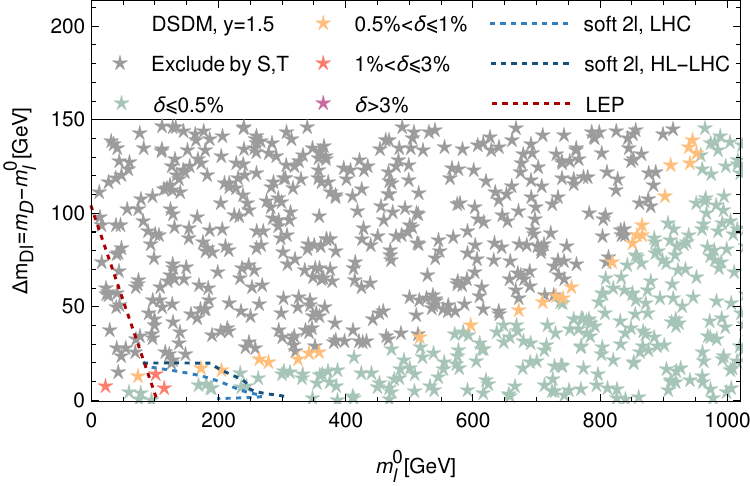}
    \caption{Parameter scan result for DSDM with $y=1$ (upper) and $y=1.5$ (lower). The LHC exclusion curves from direct searches for the new fermions are based on Ref.~\cite{Zhou:2022jgj}.}
    \label{fig:DirSDscan2}
\end{figure}

In the Dirac singlet-doublet model, the chosen set of free parameters consists of $y, m_l^0, $ and $\Delta m_{\rm Dl} = m_D-m_l^0$. Figs.~\ref{fig:DirSDscan1} and \ref{fig:DirSDscan2} display the distribution of the relative cross-section deviation $\delta$ at different values of $m_l^0$ and $\Delta m_{\rm Dl}$, with $y=1$ and $y=1.5$. A zoom-in version of Fig.\ref{fig:DirSDscan1} for small $\Delta m_{\rm Dl}$ is displayed in Fig.~\ref{fig:DirSDscan2}. In each plot, different magnitudes of the deviation $\delta$ of the expected $e^+e^-\to ZH$ cross section are represented by different colored stars. The green, yellow, red, and purple stars represent deviations of $\{0,0.5\%\}$, $\{0.5\%,1\%\}$, $\{1\%,3\%\}$, and $\{3\%,\infty\}$ respectively. 
The gray stars denote the parameter space points that are excluded by oblique parameters. 

To evaluate constraints from LHC searches, we observe that in terms of the free parameter set, the mixing angle is written as
\begin{align}
\sin^2\theta = \frac{2x^2}{2x^2+1},
\end{align}
where $x=\Delta m_{\rm Dl}/(vy)$. For large mass difference, $\sin^2\theta\approx 1$, thus the heavy charged neutral particle, $\chi^0_h$, has a dominant doublet component and becomes nearly mass degenerate with the charged particle, $\chi^\pm_D$. This scenario is similar to the bino-Higgsino system for $m_{\tilde{B}}<m_{\tilde{H}}$ (with decoupled wino) in the MSSM. If the mass difference is moderate, $\Delta m_{\rm Dl}\leq 200\gev$, $\sin^2\theta\approx 0.5$, and the cross section of $pp\to\chi_D^\pm\chi_h^0$ deviates from Higgsino production by a factor of a few. Accounting for the angle suppression, we perform the following recast 
\begin{align}
\mathcal{S}^{\rm DSDM}=\mathcal{S}^{\rm Higgsino}\times\sin^2\theta \label{eq:DSDMsig}
\end{align}
Through this recast, we derived the observed $2\sigma$ exclusion limit at the LHC from Ref.~\cite{ATLAS:2021yqv}, by implementing the HEPData from Ref.~\cite{ATLAS:2021yqv:data}, and the expected exclusion limit at the HL-LHC from Ref.~\cite{ATLAS:2018diz}. The modification of the exclusion curves due to the signal strength suppression \eqref{eq:DSDMsig} is taken into account approximately by interpolating between the 2$\sigma$ and 5$\sigma$ contours presented there. These two constraints are denoted as light and dark blue dashed lines in Fig.\ref{fig:DirSDscan1}. The red dashed line corresponds to a lower limit on chargino mass $m_D\geq 103.5$ GeV, consistent with the LEP search \cite{DELPHI:2003uqw}. For $y=1$, the most stringent limit in the large mass difference region will come from the HL-LHC search, and it can exclude the mass difference between 200 GeV and 1.1 TeV, assuming a massless $\chi_l^0$. Incorporating constraints from oblique parameters, a large fraction of the parameter space that can be probed via $\sigma_{\rm ZH}$ measurements at future colliders would already be excluded, except for small mass differences, $\Delta m_{\rm Dl} \lesssim 200$~GeV. For larger Yukawa couplings, such as $y=1.5$, the region covered by LHC searches is already excluded from LEP constraints on oblique parameters, but there is a region with large $m_D$ that would produce observable deviations $\delta > 0.5\%$ at future $e^+e^-$ colliders.

For small mass difference, $\Delta m_{\rm Dl}\leq 150$ GeV, $\sin^2\theta \ll 1$ and the heavy neutral particle, $\chi_h^0$, is predominantly a singlet. Thus the production channel involving $\chi_h^0$ is suppressed. The most important production channel is $pp\to \chi^\pm_D\chi^\mp_D$, which  is equivalent to charged Higgsino pair production in the MSSM. Using the results of the study in Ref.~\cite{Zhou:2022jgj}, we can show the expected 95\% CL exclusion contours for the LHC (assuming 100~fb$^{-1}$ at 13~TeV) and HL-LHC (3~ab$^{-1}$ at 13~TeV), which are denoted as dashed lines in Fig.\ref{fig:DirSDscan2}. Only a small region of parameter space, with $m_D \lesssim 250$~GeV and $\Delta m_{\rm Dl} \lesssim 20$~GeV can be probed by the HL-LHC. On the other hand, observable deviations $\delta > 0.5\%$ at future $e^+e^-$ colliders could be obtained in a narrow parameter region extending to $m^0_l$ values of several hundred GeV, in particular for $y>1$, which is well beyond the reach of LHC searches.

\subsection{Majorana singlet-doublet model}
\label{sec:MajSDModel}

For the Majorana singlet-doublet model, we choose $y,\,m_h^0,\,\Delta m_{\rm hl}=m_h^0-m_l^0$ as free parameters. For each viable choice of parameters $\{m_l^0,\Delta m_{\rm hl},y\}$, there are two solutions for the underlying Lagrangian parameters, $\{m_S,m_D\}$, and the mixing angle, $\sin^2\theta$. For the latter, they can be written as, in terms of the free parameters,
\begin{align}
\sin^2\theta = \frac{1}{2}\Big(1\pm \sqrt{1-\dfrac{4}{x^2}} \Big),
\end{align}
where $x=\Delta m_{\rm hl}/(vy)$. The positive solution leads to doublet-dominated $\chi_h^0$, thus we refer this solution as doublet-dominant scenario, and the other as singlet-dominant scenario. Besides, $x\geq 2$ is required to satisfy the condition $\sin^2\theta\leq 1$, which gives rise to large mass differences between $\chi_h^0$ and $\chi_l^0$, $\Delta m_{\rm hl} = \mathcal{O}(500~\gev)$. To study the LHC reach for new fermions with large mass difference, we consider SUSY collider searches with energetic leptonic and hadronic final states \cite{Liu:2020muv, ATLAS:2021yqv, CMS:2022sfi, ATLAS:2018diz,
CMS:2023qhl}.

\begin{figure}[t]
    \centering
    \includegraphics[width=0.65\textwidth]{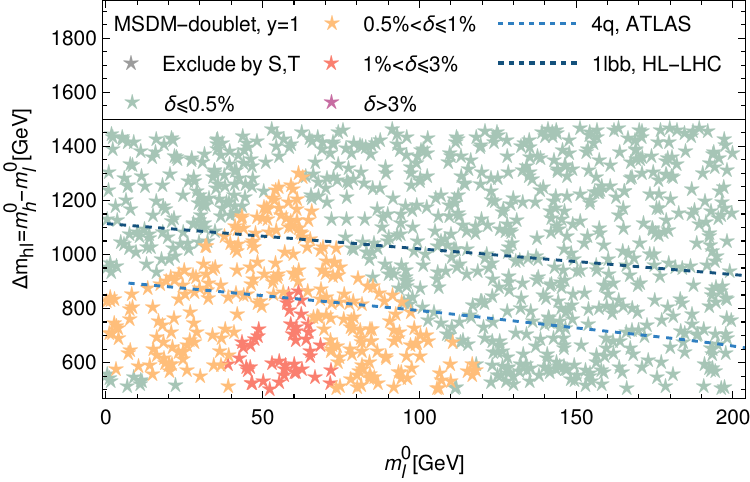}
    \\[1em]
    \includegraphics[width=0.65\textwidth]{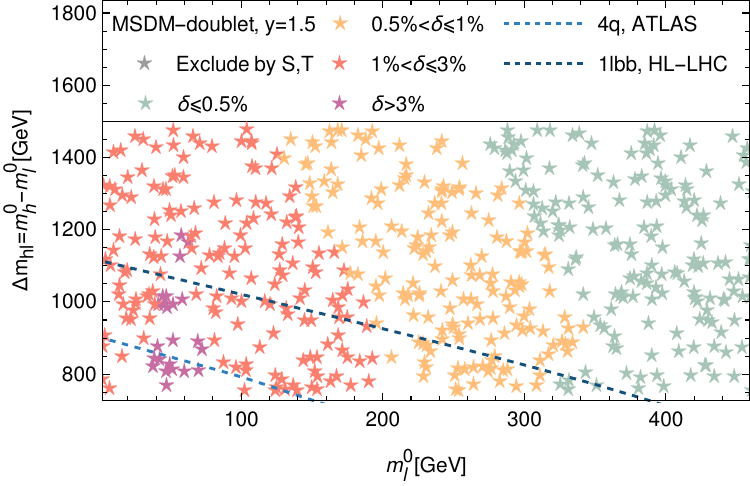}
    \caption{Parameter scan result for MSDM with $y=1$ and $y=1.5$ in doublet-dominant scenario, together with current and projected LHC constraints from Refs.~\cite{ATLAS:2021yqv} (``4q, ATLAS'') and \cite{ATLAS:2018diz}, (``1lbb, HL-LHC''), respectively.}
    \label{fig:MajSDlarge}
\end{figure}

In the doublet-dominant scenario, the dark matter candidate $\chi_l^0$ remains mostly a singlet, and all three heavy particles are predominantly doublets and exhibit nearly degenerate masses. Figure \ref{fig:MajSDlarge} displays the results of a parameter scan for $y=1$ (upper plot) and $y=1.5$ (lower plot). There are no relevant constraints from the oblique parameters in the MSDM (in particular the $T$ parameter contribution is zero in this model). 

To study the LHC phenomenology, we observe that the doublet-dominant scenario is similar to the bino-Higgsino system (with decoupled wino) in the MSSM. In fact, this correspondence would be exact for $\tan\beta=1$ and $y=g$, but any deviations from these two relations have little impact on the production cross-section, as long as the mixing angle is small, which is true in the parameter regions considered here. The most stringent constraint from current LHC data arises from chargino and neutralino searches in the fully hadronic final state \cite{ATLAS:2021yqv}, denoted as the light blue dotted line in Fig.~\ref{fig:MajSDlarge}. For small $m_l^0$, mass differences smaller than 900 GeV are excluded at a 95\% CL. The expected 95\% CL exclusion region at HL-LHC, focusing on the final state with 1 lepton and 2 b-jets\cite{ATLAS:2018diz}, places an lower limit on $\Delta m_{\rm hl}$ around 1100 GeV. For $y=1$, only a small part of the remaining parameter space points yield deviations greater than 1\%. More parameter points lead to deviations greater than 3\% for $y=1.5$. 

\begin{figure}[t]
    \centering
    \includegraphics[width=0.65\textwidth]{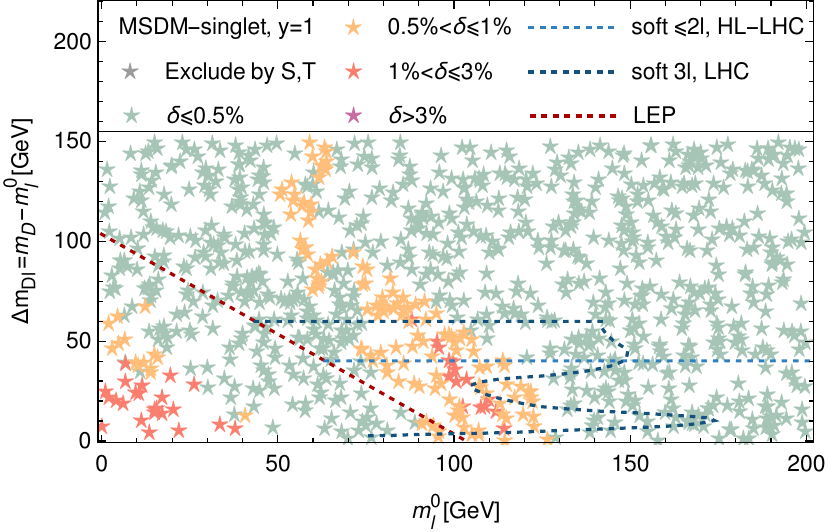}
    \caption{Parameter scan result for MSDM with $y=1$ in singlet-dominant scenario, together with direct search constraints from LEP \cite{DELPHI:2003uqw}, LHC \cite{ATLAS:2021moa} and HL-LHC \cite{Cardona:2021ebw}.}
    \label{fig:MajSDsmall}
\end{figure}

In singlet-dominant scenario, $m_h^0\approx m_S\gg m_D\approx m_l^0$. In this case, production of $\chi^0_h$ at the LHC is suppressed due to its large singlet component, thus the most important production channels are $pp\to \chi^0_l\chi^\pm_D,\,\chi^\pm_D\chi^\mp_D$. Due to the small mass difference between $\chi^\pm_D$ and $\chi^0_l$ in this scenario, we look at collider searches for compressed Higgsinos \cite{Zhou:2022jgj,Cardona:2021ebw,CMS:2018kag,ATLAS:2019lng,ATLAS:2021moa,ATLAS:2018jjf,CidVidal:2018eel}, and replace $\Delta m_{\rm hl}$ by $\Delta m_{\rm Dl} \equiv m_D-m_l^0$. The result of a parameter scan in the singlet-dominant region with fixed $y=1$ is shown in Fig.\ref{fig:MajSDsmall}. Two regions exhibit $\delta> 0.5\%$: $m_l^0\leq 50\,\mathrm{GeV}$ and $m_l^0\approx 100\,\mathrm{GeV}$. The first region leads to large relative deviations due to the threshold effects, but it is already excluded by LEP searches for charginos \cite{DELPHI:2003uqw}. In the second region, $m_h^0$ reaches its minimum value. The LHC search with soft-lepton final states \cite{ATLAS:2021moa} has excluded parameter space points within this region for which $\Delta m_{\rm Dl} \leq 60\mathrm{GeV}$ at 95\% CL. Increasing the Yukawa coupling to $1.5$ does not lead to any qualitative differences and thus is not shown.

\subsection{Dirac doublet-triplet model with triplet hypercharge \boldmath $-1$}
\label{sec:DirDT1Model}

In the Dirac doublet-triplet model with hypercharge $-1$, we consider the following free parameters: $y,m_l^0,\Delta m_{\rm ll}=m_l^\pm-m_l^0$. Similar to the MSDM case, two solutions for the mixing angle in terms of these free parameters lead to two scenarios, where the $\chi_l^\pm$ state is either mostly a doublet or mostly a triplet. We refer to these scenarios as the doublet-dominant scenario and triplet-dominant scenario, respectively.  In both scenarios, the only relevant production channel is $pp\to \chi_l^\pm \chi_l^\mp$ for two reasons\footnote{The production channel, $pp\to\chi_l^\pm\chi_l^0$, can also contribute and leads the final states with soft lepton and jet, as discussed in Ref.\cite{Schwaller:2013baa}. However, this exclusion region from this channel is weaker than $pp\to \chi_l^\pm \chi_l^\mp$, and thus not included.}: (a) $\chi_l^\pm$ is the lightest unstable particle; (b) all the other heavy particles are much heavier and effectively decouple. To illustrate the latter explicitly, we can express the masses of decoupled heavy particles in terms of $\Delta m_{ll}$,
\begin{align}
&\frac{m_h^{0}-m_l^0}{\Delta m_{ll}}= 1 + \frac{1}{4x^2} ~,~\notag\\
&\frac{m_h^{\pm}-m_l^0}{\Delta m_{ll}}=\frac{1}{4x^2}~,~\notag\\
&\frac{m^{--}-m_l^0}{\Delta m_{ll}} =  \frac{1}{2}+\frac{1}{8x^2}+\frac{1}{8}\sqrt{16-\frac{24}{x^2}+\frac{1}{x^4}} ,
\end{align}
where $x=\Delta m_{ll}/(vy)$. To ensure all parameters are real, the condition $x^2\leq (3/4-1/\sqrt{2})=0.043$, is imposed\footnote{$x^2\geq (3/4+1/\sqrt{2})$ also leads to real parameters, but it is not compatible with the mass ordering $m_l^0 < m_l^\pm < m_h^{0,\pm}$.}, which leads to relatively heavy $m_h^{0,\pm}$ and $m^{--}$. For example, for $y=2$ and $\Delta m_{ll}<50$~GeV, one has $m_h^0>1.25$~TeV. Thus the production cross section involving those particles can be ignored. 

\begin{figure}[h!]
    \centering
    \includegraphics[width=0.65\textwidth]{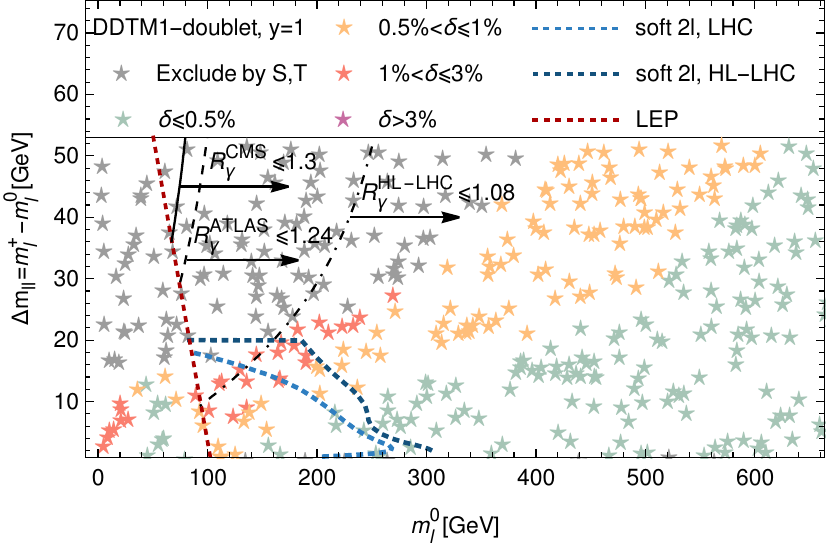}
    \\[1em]
    \includegraphics[width=0.65\textwidth]{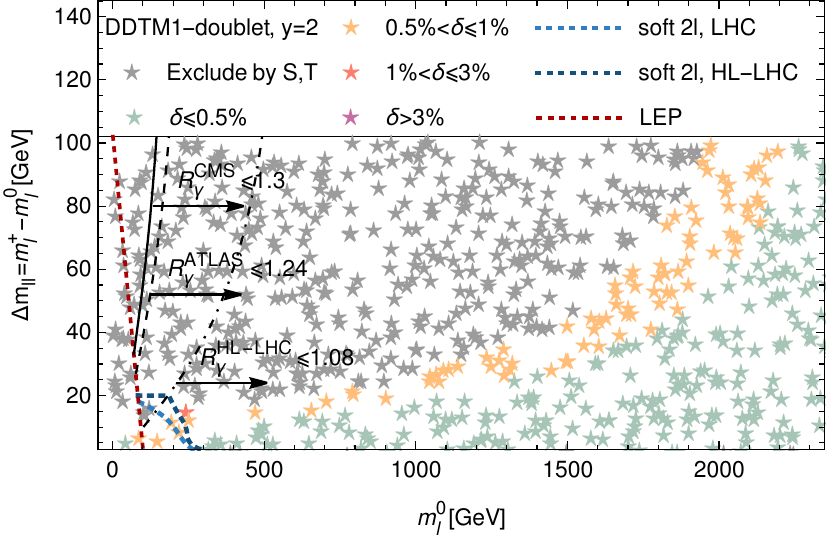}
    \caption{Parameter scan result for DDTM1 with $y=1$(upper) and $y=2$(lower) in the doublet-dominant scenario. The light blue and dark blue dashed line represent the expected 95\% C.L. reach of 100$\text{fb}^{-1}$ and 3$\text{ab}^{-1}$ LHC13 from Ref.\cite{Zhou:2022jgj}, denoted as "soft 2l,LHC" and "soft 2l, HL-LHC" respectively. The constraint of LEP \cite{DELPHI:2003uqw} is the dark red dashed line. The black lines denote the upper limits of $R_\gamma$ from the CMS \cite{CMS:2021kom}, ATLAS \cite{ATLAS:2022tnm} as well as HL-LHC\cite{Cepeda:2019klc}, respectively.}
    \label{fig:DirDTdoublet}
\end{figure}

\begin{figure}[h!]
    \centering
    \includegraphics[width=0.65\textwidth]{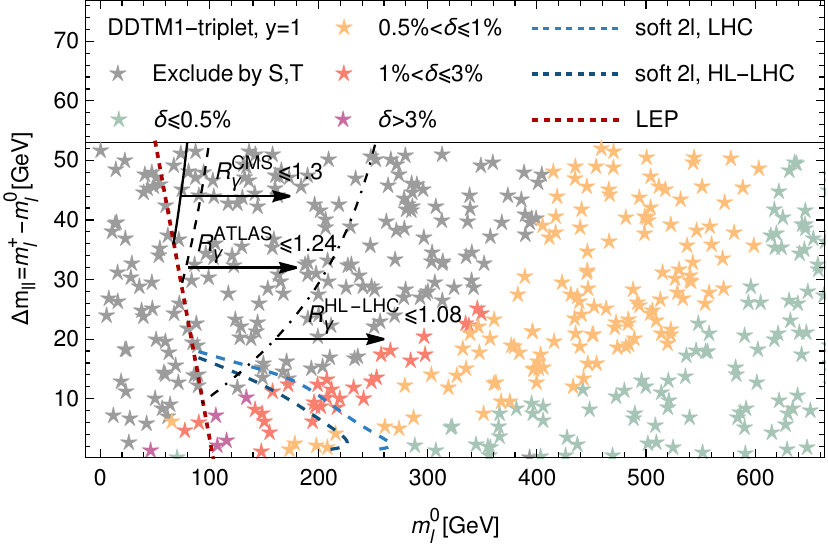}
    \\[1em]
    \includegraphics[width=0.65\textwidth]{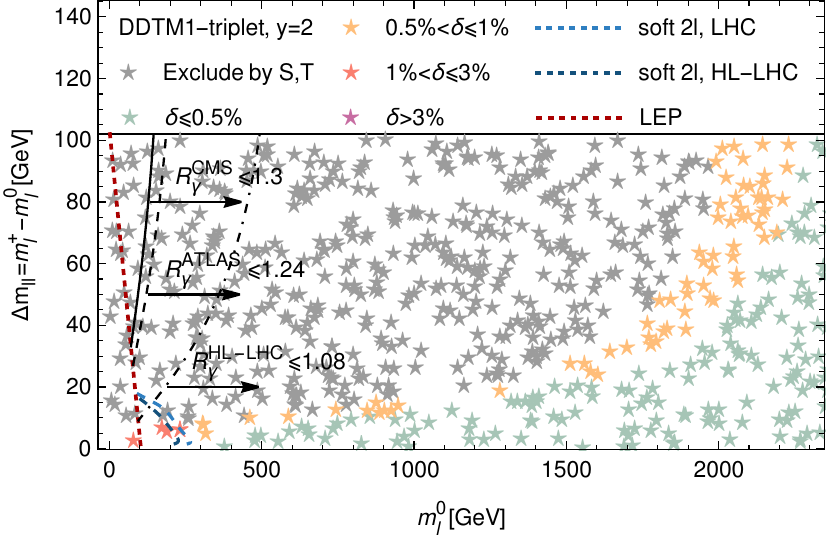}
    \caption{Parameter scan result for DDTM1 with $y=1$(upper) and $y=2$(lower) in the triplet-dominant scenario. The light blue and dark blue dashed line represent the expected 95\% C.L. reach of 100$\text{fb}^{-1}$ and 3$\text{ab}^{-1}$ LHC13 from Ref.\cite{Zhou:2022jgj}, denoted as "soft 2l,LHC" and "soft 2l, HL-LHC" respectively. These constraints are rescaled according to Eq.~\ref{eq:recastDouTri}. The constraint of LEP \cite{DELPHI:2003uqw} is the dark red dashed line. The black lines denote the upper limits of $R_\gamma$ from the CMS \cite{CMS:2021kom}, ATLAS \cite{ATLAS:2022tnm} as well as HL-LHC\cite{Cepeda:2019klc}, respectively.}
    \label{fig:DirDTtriplet}
\end{figure}

The distributions of $\delta$ in the doublet- and triplet-dominate scenario are shown in Fig.~\ref{fig:DirDTdoublet} and Fig.~\ref{fig:DirDTtriplet} respectively. In both scenarios, the Yukawa coupling is considered with values 1 and 2. For same Yukawa coupling, $\delta$ exhibits similar distributions in both scenarios. The constraint from the Higgs to di-photon decay branching ratio is denoted as the black solid, dashed and dot-dashed lines, which represent the upper limits of $R_\gamma$ from the CMS \cite{CMS:2021kom}, ATLAS \cite{ATLAS:2022tnm} as well as HL-LHC\cite{Cepeda:2019klc}, respectively. The whole parameter region shown in Fig.~\ref{fig:DirDTdoublet} and Fig.~\ref{fig:DirDTtriplet} satisfies the lower bounds on $R_\gamma$, thus they are not visible in the figures. 

For the case of $y=1$, the oblique parameters and the $R_\gamma$ constraint exclude the region of large mass differences for small $m_l^0$. The majority of the surviving parameter points yield $\delta> 0.5\%$, while only a few points result in $\delta> 1\%$. As the Yukawa coupling increases, a greater number of parameter points are excluded by the oblique parameters, but there is a slice of unconstrained parameter space with $\delta >0.5\%$ extending to multi-TeV values of $m_l^0$. 

For a doublet $\chi_l^\pm$, the production channel, $pp\to\chi_l^\pm\chi_l^\mp$, is equivalent to charged Higgsino pair production. Besides, the condition $x^2\leq 0.043$ causes $\Delta m_{ll} \lesssim \mathcal{O}(50\text{ GeV})$ for a $\mathcal{O}(1)$ Yukawa coupling. As explained in the previous subsection, scenarios with such small mass differences can be best searched for by using the hard jet plus soft leptons signature at the LHC. The expected 95\% CL reach of LHC13 with 100~$\text{fb}^{-1}$ and 3~$\text{ab}^{-1}$ from the analysis in Ref.\cite{Zhou:2022jgj} is shown by the light and dark blue dotted lines in Fig.~\ref{fig:DirDTdoublet} respectively. The exclusion contour from chargino searches at LEP \cite{DELPHI:2003uqw} (red dotted line) is also shown in the figure. These bounds are also applied to the triplet-dominant scenario. To account for the differences of the production cross section between the doublet- and triplet-dominant scenario, the following recast was performed:
\begin{align}
\mathcal{S}^{\text{Triplet}} = \mathcal{S}^{\text{Doublet}} \times \frac{\sigma^{\text{Triblet}}(pp\to \chi_l^\pm\chi_l^\mp)}{\sigma^{\text{Doublet}}(pp\to \chi_l^\pm\chi_l^\mp)} \approx \mathcal{S}^{\text{Doublet}} \times \frac{1}{5} \label{eq:recastDouTri}
\end{align}
where $\mathcal{S}$ denotes the signal significance and ``Triplet(Doublet)'' stands for the triplet(doublet)-dominant scenario. This equation implies that the $95\%$ CL exclusion contour for a triplet $\chi_l^\pm$ corresponds to $10\sigma$ exclusion contour in the doublet case, which is obtained by extrapolating Table~4 of Ref.~\cite{Zhou:2022jgj}, resulting in the contours shown in Fig.~\ref{fig:DirDTtriplet}. As is evident from Fig.~\ref{fig:DirDTdoublet} and Fig.~\ref{fig:DirDTtriplet}, the exclusion contour in triplet-dominant scenario is smaller than the doublet one due to the suppression of the production cross section.

As can be seen from the yellow and red points in Fig.~\ref{fig:DirDTdoublet}, precision measurements of the $ZH$ cross-section at $e^+e^-$ colliders can probe significant parameter regions of that are beyond the reach of the LHC, namely for $\Delta m_{ll} > 20$~GeV or values of $m_l^0$ of several hundred GeV. For larger Yukawa couplings ($y=2$), the interesting parameter region is shifted to larger masses, since the constraints from oblique parameters are stronger, but there is still substantial territory where it is possible to have deviations $\delta > 0.5\%$ for 200~GeV $\gtrsim m_l^0 \gtrsim$ 2 TeV. A similar qualitative behaviour is observed for the triplet-dominant scenario, Fig.~\ref{fig:DirDTtriplet}. In particular, due to the suppressed LHC production cross-section in this scenario, the LHC can only cover a small part of the parameter region where $\delta > 0.5\%$.

\subsection{Dirac doublet-triplet model  with triplet hypercharge 0}
\label{sec:DirDT0Model}

For Dirac doublet-triplet model with zero hypercharge, the free parameter set is chosen to be $\{y,\,m_l^0,\,\Delta m_{\rm hl}=m_h^0-m_l^0\}$. Similar to the MSDM, the mixing angles have two different values for each choice of free parameters $\{y,m_{\rm hl}\}$, and are written as
\begin{align}
\sin^2\theta' &= \frac{1}{2}\Big(1\pm \sqrt{1-x^{-2}}\Big), \\    
\sin^2\theta &= \frac{1}{2}\Big(1\pm \frac{\sqrt{1-x^{-2}}}{\sqrt{1+x^{-2}}}\Big),    
\end{align}
where $x=\Delta m_{\rm hl}/(vy)$. We refer to the positive solution as the "doublet-dominated scenario," in which $\chi_h^{0,\pm}$ are mainly doublets, and the other solution stands for the "triplet-dominated scenario."

\begin{figure}[h!]
    \centering
    \includegraphics[width=1\textwidth]{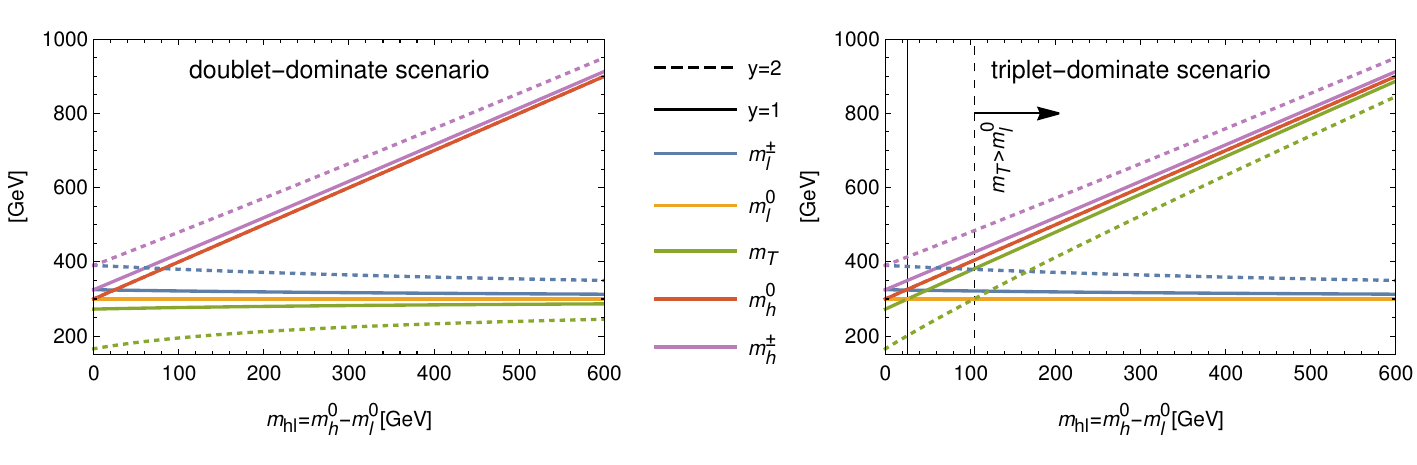}
    \vspace{-2em}
    \caption{The mass distribution of five particles in the Dirac doublet-triplet model with hypercharge $Y=0$ (DDTM0), as functions of $\Delta m_{\rm hl}$ in the doublet-dominate scenario(left) and triplet-dominate scenario(right). The Yukawa coupling is chosen to be: $y=1$ (solid) and $y=2$ (dashed), and $m_l^0=300$ GeV in both plots. }
    \label{fig:DDTM0massDouTri}
\end{figure}

In Fig.~\ref{fig:DDTM0massDouTri}, the mass distributions of the five particles are shown as a function of $\Delta m_{\rm hl}$ for two choices of the Yukawa coupling, $y = 1$ and $y = 2$, while $m_l^0$ is set to be 300 GeV. In the doublet-dominated scenario, the lightest particle is $\chi_m^\pm$, which is not charge neutral and, if it is stable, is inconsistent with cosmological constraints. However, in the triplet-dominated scenario, $\chi_l^0$ can be the lightest particle if the mass difference satisfies $\Delta m_{\rm hl} > v^2y^2/(8m_l^0)$. It is also evident from the figure that $m_l^0 \approx m_l^\pm$ and $m_T \approx m_h^0 \approx m_h^\pm$. This is related to the fact that in this scenario the mixing angles are  $\cos\theta \approx \cos\theta' \approx 1$, which indicates that $\chi_l^{0,\pm}$ are doublet-dominant states, while $\chi_h^{0,\pm}$ are predominantly triplets, \emph{i.e.} their masses are close to $m_T$.

\begin{figure}[tp]
    \centering
    \includegraphics[width=0.63\textwidth]{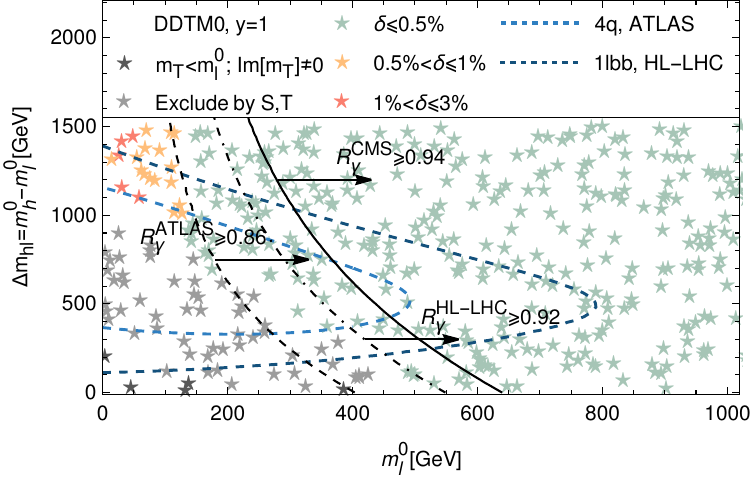}\\[1em]
    \includegraphics[width=0.63\textwidth]{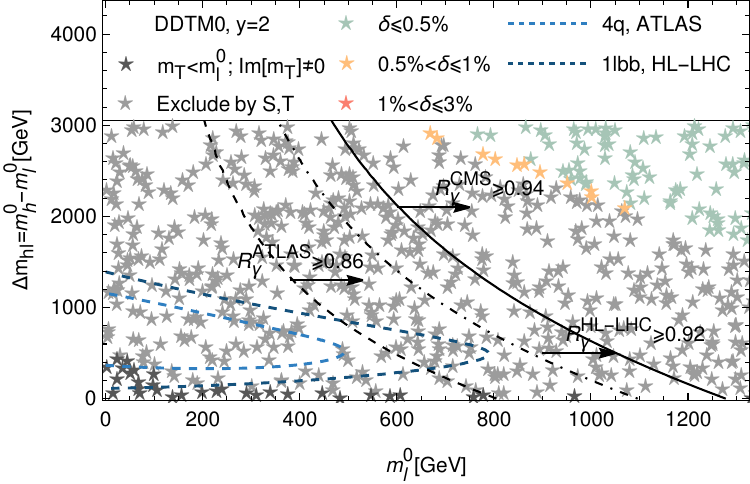}\\[1em]
    \includegraphics[width=0.63\textwidth]{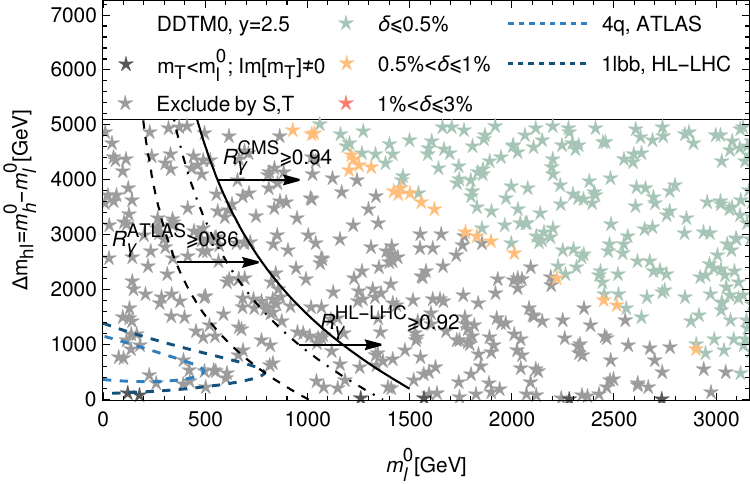}    
    \caption{Parameter scan result for DDTM0 with $y=1$, $y=2$ and $y=2.5$ in the large mass difference region, together with current and projected LHC constraints from Refs.~\cite{ATLAS:2021yqv} (``4q, ATLAS'') and \cite{ATLAS:2018diz}, (``1lbb, HL-LHC''), as well as the constraint from branching fraction of the Higgs boson to di-photons from \cite{ATLAS:2022tnm,CMS:2021kom, Cepeda:2019klc}.}
    \label{fig:scanDDTM}
\end{figure}

Fig.~\ref{fig:scanDDTM} displays $\delta$ in the triplet-dominant scenario at different values of $m_l^0$ and $\Delta m_{\rm hl}$, and the Yukawa coupling is chosen to be $y=1(2,2.5)$ in the upper(middle,lower) plot. In the figure, a new type of point is introduced and represented by dark gray stars. These points correspond to parameter choices that result in complex masses or $m_T < m_l^0$, and they are excluded since they are unphysical. This exclusion removes a few points in the region of small mass differences. Additionally, more points in the small mass difference region are excluded due to constraints from oblique parameters.

Viable parameter space points with $\delta> 0.5\%$ are limited and they are concentrated in the region with large mass difference, which can be best constrained through collider searches using energetic leptons and hadronic jets in the final states. The relevant channels include the production of charged states pairs, $pp\to \chi_h^\pm\chi_h^\mp,\chi_m^\pm\chi_m^\mp$,\footnote{The production channel $pp\to \chi_l^\pm\chi_l^\mp$ is not taken into account since $m_l^\pm-m_l^0=\mathcal{O}(20~{\rm GeV})$.}, as well as the production of charged-neutral pairs, $pp\to\chi_h^0\chi_h^\pm,\chi_h^0\chi_m^\pm$.\footnote{The production of neutral pairs, $pp\to \chi_h^0\chi_h^0$, can also contribute. However, the channel is significantly suppressed by the factor $\sin^2\theta\approx 0$.} The cross section summing over all channels are approximately twice the cross section of wino-pair production, $pp\to \Tilde{W}^\pm\Tilde{W}^\mp+\Tilde{W}^0\Tilde{W}^\pm$. Consequently, the 95\% CL contour in this model corresponds to the $1\sigma$ exclusion region in the search for Wino-pair production. We use the results from Refs.~\cite{ATLAS:2021yqv,ATLAS:2018diz}. The LHC bound is obtained by extrapolating the bounds for wino and Higgsino pair production (which has a cross-section of roughly half the wino pair production cross-section), with the help of HepData \cite{ATLAS:2021yqv:data,ATLAS:2021yqv:dataWino}. The exclusion contour at the HL-LHC is obtained by extrapolating the $2\sigma$ and $5\sigma$ bounds presented there.

The resulting estimated exclusion contours are incorporated into Fig.~\ref{fig:scanDDTM}. Among these searches, the most stringent constraint arises from the HL-LHC projection, which excludes mass differences up to 1.4 TeV assuming a massless dark matter candidate. Combining the constraint from the Higgs diphoton decay branching fraction, all parameter points with $\delta > 0.5\%$ within the region depicted in Fig.~\ref{fig:scanDDTM} are excluded for $y=1$, while surviving points emerge for $\Delta m_{\rm hl}\gtrsim 4.6$ TeV. There exist a few surviving parameter points with $\delta > 0.5\%$ in the case of $y=2$, with TeV-scale fermion masses and large mass differences. For even larger Yukawa couplings, such as $y=2.5$ shown in Fig.\ref{fig:scanDDTM} (bottom), the entire parameter region with $m_l^0\leq 1~{\rm TeV}$ is excluded by the oblique parameters, but observable deviations $\delta$ of the $e^+e^- \to ZH$ cross-section are found for fermion masses beyond 1.5~TeV. This is well beyond the reach of the (HL-)LHC, whereas this region can be explored at a 100 TeV hadron collider, as analyzed in Ref.\cite{Gori:2014oua}. The precision measurements of the cross section for $e^+e^-\to ZH$ can offer complementary information to the direct searches conducted at the 100 TeV hadron colliders. 

\subsection{Majorana doublet-triplet model}
\label{sec:MajDTModel}

For the Majorana doublet-triplet model, $y$, $m_l^0$ and $\Delta m_{\rm hl}=m_h^0-m_l^0$ are chosen as free parameters. Similar to the Majorana singlet-doublet model, for each choice of free parameters $\{y,\Delta m_{\rm hl}\}$, the mixing angle has two different possible values
\begin{align}
\sin^2\theta = \frac{1}{2}\Big(1\pm \sqrt{1-\dfrac{2}{x^2}} \Big), 
\end{align}
where $x=\Delta m_{\rm hl}/(vy)\geq \sqrt{2}$ to ensure all parameters are real. Following what we did in the Majorana singlet-doublet model, we will refer the positive solution as "doublet-dominant" scenario, under which $\chi_h^{0,+}$ is mostly a doublet with a small triplet admixture, and the second as "triplet-dominant" scenario. Besides, $x\geq \sqrt{2}$,  which corresponds to a mass difference of the order $\Delta m_{\rm hl}=\mathcal{O}(400~\text{GeV})$, must be satisfied to ensure all parameters are real. As discussed in Sec.~\ref{sec:MajSDModel}, collider searches for signatures with energetic leptons and hadronic jets can put stringent bounds on dark sector fermions with large mass differences. We consider the searches reported in Refs.~\cite{ATLAS:2021yqv, CMS:2022sfi, CMS:2023qhl} and projections from Refs.~\cite{Liu:2020muv,ATLAS:2018diz}. 

In the doublet-dominant scenario, three heavy particles are all doublets and nearly mass degenerate, while two light particles are triplets, which is similar to the wino-Higgsino scenario for $m_{\tilde{W}}<m_{\tilde{H}}$ (with decoupled bino) in MSSM. In the wino-Higgsino scenario, it is typically assumed that the three Higgsino components are mass degenerate, namely $m_{\tilde{\chi}_1^\pm} = m_{\tilde{\chi}_1^0} = m_{\tilde{\chi}_2^0}$. However, in our model, the mass ordering is such that $m_m^0 < m_h^0 = m_h^\pm$. Despite this difference, when considering the effects of mixing angles and mass differences, the modifications to the cross section are found to be less than 10\%. Consequently, the mass difference is adjusted by approximately $\mathcal{O}(10~\text{GeV})$, which is small and causes no qualitative difference. Thus the results in wino-Higgsino scenario from Ref.~\cite{ATLAS:2021yqv} can be directly implemented. Moreover, it is worth mentioning that the exclusion limits for bino-Higgsino scenario for $m_{\tilde{B}}<m_{\tilde{H}}$ \cite{Liu:2020muv,CMS:2022sfi, ATLAS:2018diz,CMS:2023qhl} can also be utilized, which is due to the observation that exclusion limits for the wino-Higgsino scenario are almost the same as for the bino-Higgsino scenario \cite{ATLAS:2021yqv}. 

In the triplet-dominant scenario, the dark matter production channels include $pp\to \chi_h^\pm\chi_h^\mp,\chi_h^0\chi_h^\pm,\chi_m^0\chi_h^{0,\pm}$, where $\chi_h^{0,\pm}$ are mass degenerate triplet-dominant states, and $\chi_m^0$ is pure doublet. Production channels involving $\chi_m^0$ are suppressed by $\sin^2\theta\ll 1$. Thus the dominant production channels are $pp\to \chi_h^\pm\chi_h^\mp,\chi_h^0\chi_h^\pm$, which is equivalent to the Higgsino-wino scenario ($m_{\tilde{H}}<m_{\tilde{W}}$). Similar to above, according to the analysis in Ref.~\cite{ATLAS:2021yqv}, the exclusion limits for the Higgsino-wino scenario are very similar to those of the bino-wino scenario ($m_{\tilde{B}}<m_{\tilde{W}}$). Therefore, we also incorporate the exclusion contours obtained from studies on the bino-wino scenario \cite{CMS:2022sfi, ATLAS:2018diz, CMS:2023qhl}.

\begin{figure}[h!]
    \centering
    \includegraphics[width=0.65\textwidth]{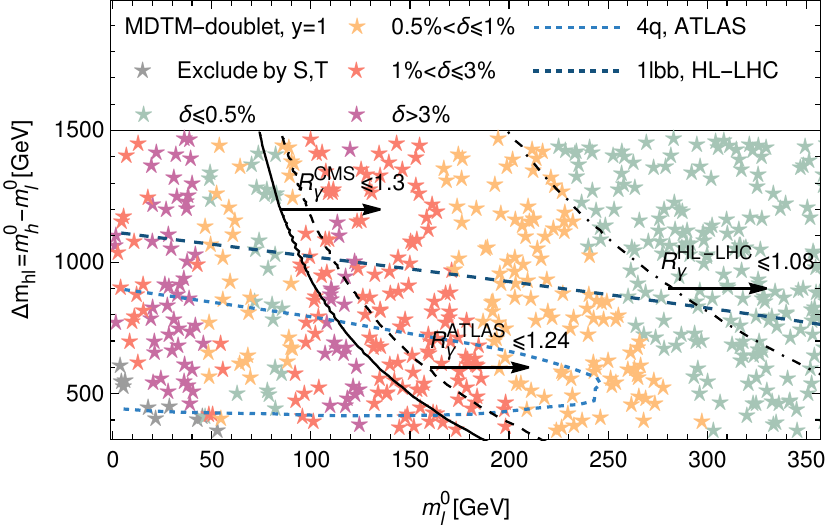}
    \\[1em]
    \includegraphics[width=0.65\textwidth]{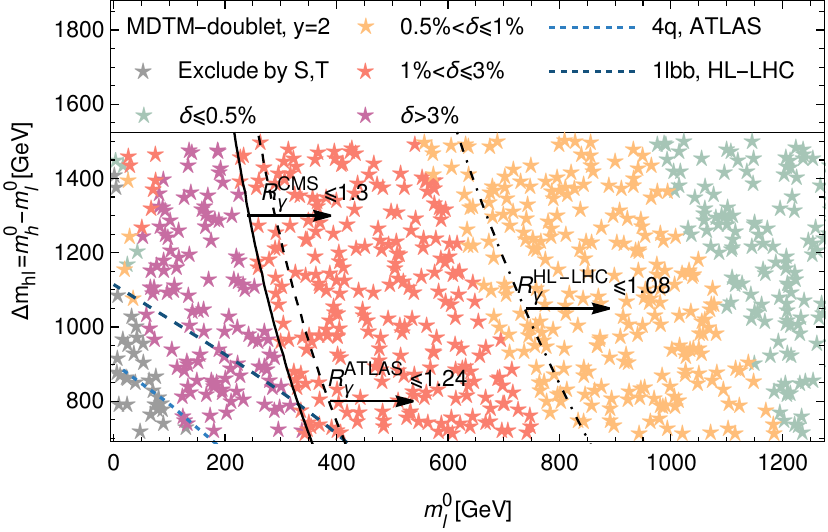}
    \caption{Parameter scan result for MDTM with $y=1$ and $y=2$ in doublet-dominant scenario, together with current and projected LHC constraints from Refs.~\cite{ATLAS:2021yqv} (``4q, ATLAS'') and \cite{ATLAS:2018diz}, (``1lbb, HL-LHC''), as well as the constraint from branching fraction of the Higgs boson to di-photons from \cite{ATLAS:2022tnm,
CMS:2021kom,Cepeda:2019klc}. }
    \label{fig:MajDTdoublet}
\end{figure}

\begin{figure}[h!]
    \centering
    \includegraphics[width=0.65\textwidth]{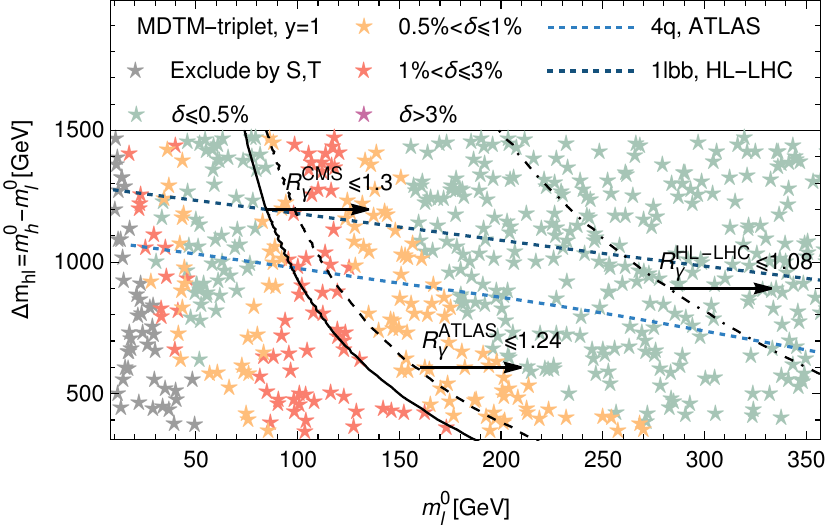}
    \\[1em]
    \includegraphics[width=0.65\textwidth]{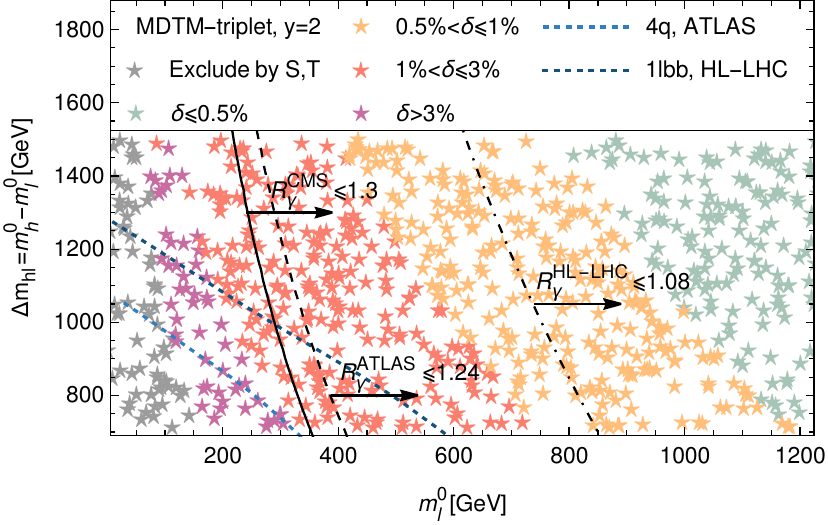}
    \caption{Parameter scan result for MDTM with $y=1$ and $y=2$ in triplet-dominant scenario, together with current and projected LHC constraints from Refs.~\cite{ATLAS:2021yqv} (``4q, ATLAS'') and \cite{ATLAS:2018diz}, (``1lbb, HL-LHC''), as well as the constraint from branching fraction of the Higgs boson to di-photons from \cite{ATLAS:2022tnm,
CMS:2021kom,Cepeda:2019klc}. }
    \label{fig:MajDTtriplet}
\end{figure}

Fig.\ref{fig:MajDTdoublet} and Fig.\ref{fig:MajDTtriplet} display the scan results in the doublet- and triplet-dominant scenario, respectively, together with the constraints from oblique parameters, branching fraction of the Higgs boson to di-photons, as well as the collider searches and HL-LHC projections. The oblique parameters exclude small values of $m_l^0$, up to about 50~GeV for $y=1$ and about 150~GeV for $y=2$. The constraint from Higgs to di-photon decay branching ratio is denoted as the black solid, dashed and dot-dashed lines, which represent the upper limits of $R_\gamma$ from the CMS \cite{CMS:2021kom}, ATLAS \cite{ATLAS:2022tnm} as well as HL-LHC\cite{Cepeda:2019klc}, respectively. For $y=2$, the constraint from the Higgs branching fraction into photons are becomes stronger, excluding masses of $\chi_l^0$ up to 700--850~GeV. At the same time, precision measurements of $e^+e^- \to ZH$ can probe a larger region of parameter space for the larger Yukawa coupling $y=2$, with deviations $\delta > 0.5\%$ possible for $\mathcal{O}$(TeV) masses of the new fermions. In the triplet-dominant scenario, the exclusion contours from direct LHC searches are more stringent due to the higher pair production cross section of winos compared to Higgsinos. The expected exclusion contour for the HL-LHC significantly extends beyond the current limits, excluding triplet masses up to 1.3 TeV and doublet masses up to 1.1 TeV, assuming a massless $\chi_l^0$. For $y=1$, this would exclude most of the parameter points that could be accessible via $ZH$ production at $e^+e^-$ collider, but there is still an interesting region for $m_l^0 \lesssim 200$~GeV and large mass differences $\Delta m_{\rm hl}$ beyond 1~TeV. This region can be covered at the HL-LHC through the precision measurement on $R_\gamma$. For larger Yukawa coupling, $y=2$, the $e^+e^- \to ZH$ cross-section is sensitive to larger regions of parameter space beyond the reach of the HL-LHC, extending to multi-TeV values for $m_h^0$.

\clearpage

\begin{figure}[t]
    \centering
    \includegraphics[width=0.65\textwidth]{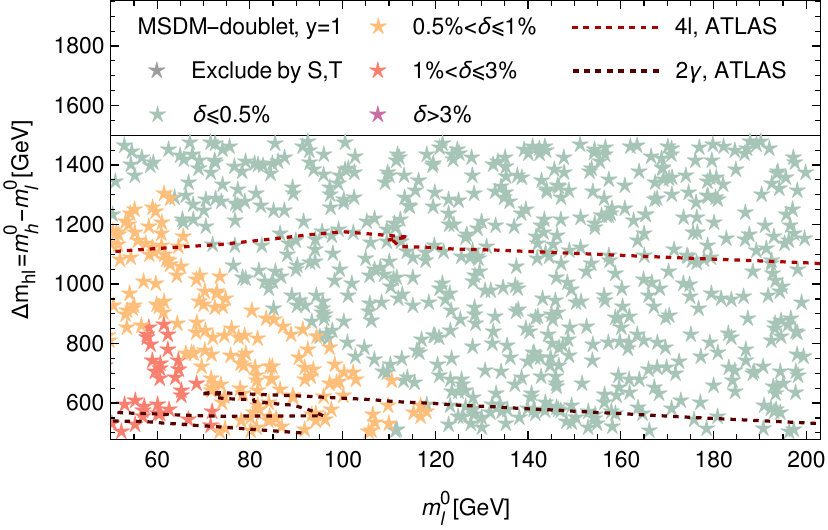}
    \\[1em]
    \includegraphics[width=0.65\textwidth]{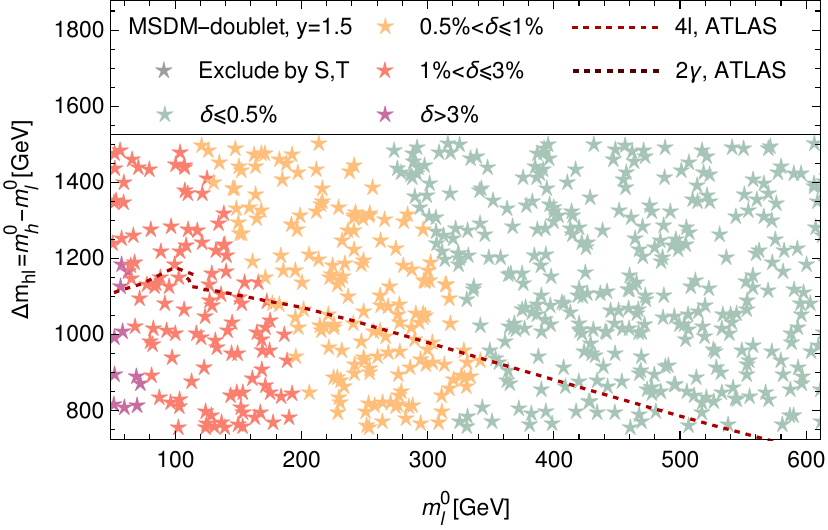}
    \caption{Parameter scan result for the MSDM with $y=1$(upper) and $y=1.5$(lower) in the doublet-dominant scenario. The solid curve indicates the exclusion for leptonic decays of $\chi^0_l$ using multi-lepton RPV SUSY searches at LHC13 \cite{ATLAS:2021yyr} (``4l, ATLAS''). The dashed curve displays the exclusion for $\chi^0_l$ decays into final states with photons, using searches for gauge-mediated SUSY at LHC13 \cite{ATLAS:2018nud} (``$2\gamma$, ATLAS''). In both cases the region below the curves are excluded.}
    \label{fig:MSDMdec}
\end{figure}

\begin{figure}[t]
    \centering
    \includegraphics[width=0.65\textwidth]{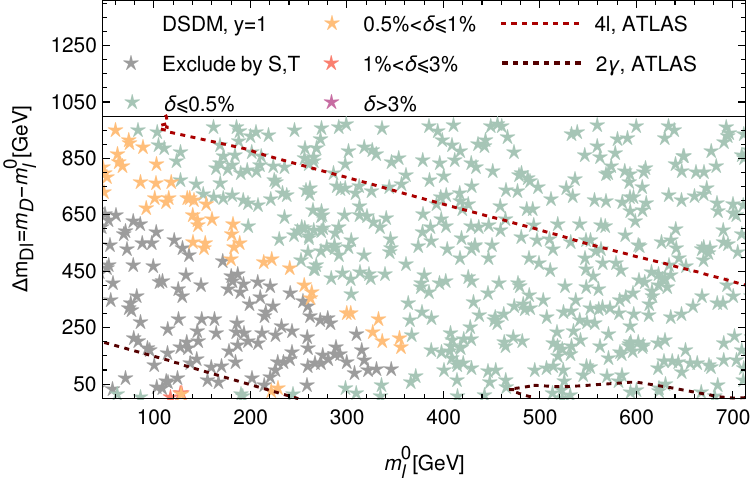}
    \\[1em]
    \includegraphics[width=0.65\textwidth]{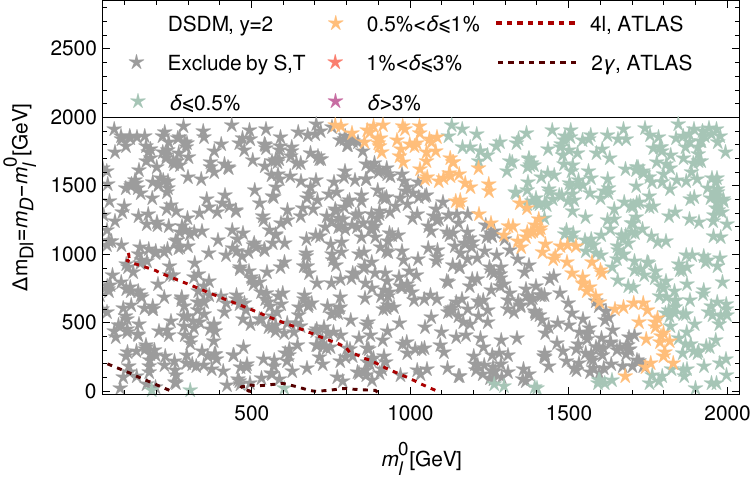}
    \caption{Parameter scan result for the DSDM with $y=1$(upper) and $y=2$(lower) in the doublet-dominant scenario. See caption of Fig.~\ref{fig:MSDMdec} for the definition of the curves.}
    \label{fig:DSDMdec}
\end{figure}

\section{Results for non-minimal scenarios with decaying dark sector fermions} \label{sec:result2}

In the previous section we considered models where the lightest new fermion is assumed to be stable, leading to missing energy signatures at the LHC. In general, however, it is also possible that this particle decays into other BSM particles and/or SM particles through additional dark sector interactions. For concreteness, let us assume that any additional dark sector states do not have significant couplings to the SM Higgs and gauge bosons, which implies that the corrections to $\sigma_{\rm ZH}$ and the S and T parameters are the same as in the previous section. However, the observable signatures at the LHC are modified due to the extended decay chains. In particular, additional visible decay products from the decay of $\chi^0_l$ can help to discriminate the BSM signal from SM background and thus improve the LHC sensitivity.

While the full range of possibilities for the $\chi^0_l$ decays is too extensive to be comprehensively explored here, we consider two scenarios that are particularly interesting for the LHC phenomenology: $\chi^0_l$ decays into final states with leptons and into final states with photons.
The former is similar to SUSY scenarios with an R-parity violating $LLE$ coupling in the superpotential, whereas the latter is similar to gauge-mediated supersymmetry, where the bino decays into a photon and a gravitino. Therefore we derive LHC limits by recasting corresponding SUSY searches. Specifically, for the R-parity violating (RPV) leptonic decay, we use cross-section limits from the study of Ref.~\cite{ATLAS:2021yyr} provided through HEPData \cite{ATLAS:2021yyr:data}. For the photonic signatures, we employ the study of Ref.~\cite{ATLAS:2018nud}, which also has a HEPData repository, where the relevant cross-section limits can be found at Ref.~\cite{ATLAS:2018nud:data}.

\medskip

For the singlet-doublet models, MSDM and DSDM, we here focus on the doublet-dominated scenarios, where the interesting parameter region is much larger than for the singlet-dominated scenarios. As mentioned in the previous section, pair production of new doublet fermions is similar to Higgsino pair production in the MSSM. Therefore, the LHC constraints can be obtained by comparing the ATLAS cross-section limits to the Higgsino pair production cross-section \cite{gaugino} from the LHC SUSY Cross Section Working Group (see Ref.~\cite{Fuks:2012qx,Fuks:2013vua} for the cross-section calculations). The results are shown in Figs.~\ref{fig:MSDMdec} and \ref{fig:DSDMdec}, respectively.

For the scenario of the leptonically decaying $\chi^0_l$, the constraints from LHC13 are fairly strong. They exclude almost all the parameter points with $\delta > 0.5\%$ for moderate values of the Yukawa coupling, $y=1$, in both the MSDM and the DSDM. For larger Yukawa couplings, on the other hand, there are significant regions of parameters space where an observable deviation of the $\sigma_{\rm ZH}$ could be obtained that are not excluded by LHC13 data. The HL-LHC will lead to stronger constraints, but it is clear that is cannot cover the entirety of these parameter regions.

For the scenario of $\chi^0_l$ decaying into photon final-states, on the other hand, the constraints from LHC13 are much weaker. Even for $y=1$, most of the parameter space with $\delta>0.5\%$ is not constrained by existing ATLAS results, and this is unlikely to change dramatically for the HL-LHC. Therefore, for the photonic decays of $\chi^0_l$, future $e^+e^-$ colliders are a promising tool for probing the MSDM and DSDM models.

\medskip

Now let us consider the doublet-triplet models. For the doublet-dominated (triplet-dominated) scenarios, the LHC exclusion curves are obtained by comparing the cross-section limits from the ATLAS HEPData repositories to Higgsino (wino) pair production cross-sections. As before, we use computed cross-section values from the LHC SUSY Cross Section Working Group \cite{gaugino} for that purpose. 

\begin{figure}[t]
    \centering
    \includegraphics[width=0.65\textwidth]{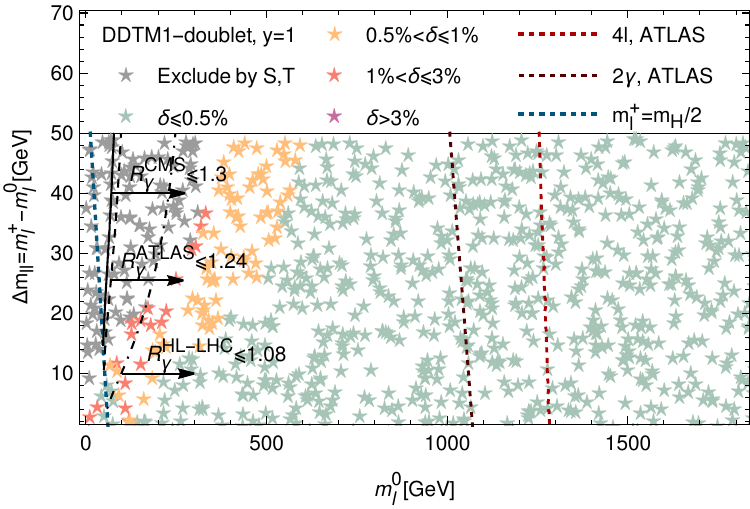}
    \\[1em]
    \includegraphics[width=0.65\textwidth]{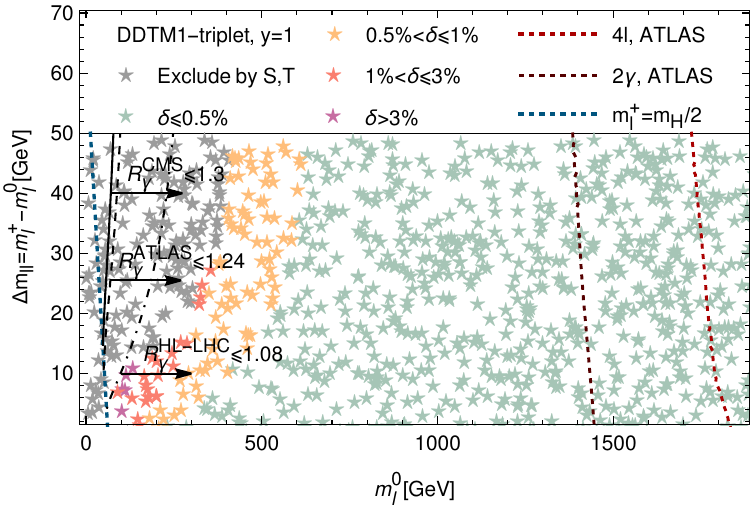}
    \caption{Parameter scan result for the DDTM1 with $y=1$ doublet-dominant (upper) and triplet-dominant (lower) scenario. See caption of Fig.~\ref{fig:MSDMdec} for the definition of the curves. The black lines denote the upper limits of $R_\gamma$ from the CMS \cite{CMS:2021kom}, ATLAS \cite{ATLAS:2022tnm} as well as HL-LHC\cite{Cepeda:2019klc}, respectively.}
    \label{fig:DDTM1dec}
\end{figure}

Fig.~\ref{fig:DDTM1dec} shows the results in the DDTM1 with Yukawa couplng $y=1$. As evident from the figure, the entire parameter space that can lead to visible deviations in $ZH$ production at $e^+e^-$ colliders is already ruled out by existing LHC data, both for the leptonic and for photonic $\chi^0_l$ decays. Compared to the scenarios with stable $\chi^0_l$ (see Figs.~\ref{fig:DirDTdoublet}, \ref{fig:DirDTtriplet}), the additional leptons or photons from $\chi^0_l$ decay help to produce much stronger LHC constraints in this model, which is characterized by small mass differences between the lighter part of the new fermion spectrum. The same conclusion holds for other values of the Yuakwa coupling in the DDTM1. 

Since the viable parameter space for the DDTM0 is strongly constrained by bounds from the oblique parameters and the Higgs branching factor $R_\gamma$, we do not explore this model here further.

\begin{figure}[t]
    \centering
    \includegraphics[width=0.65\textwidth]{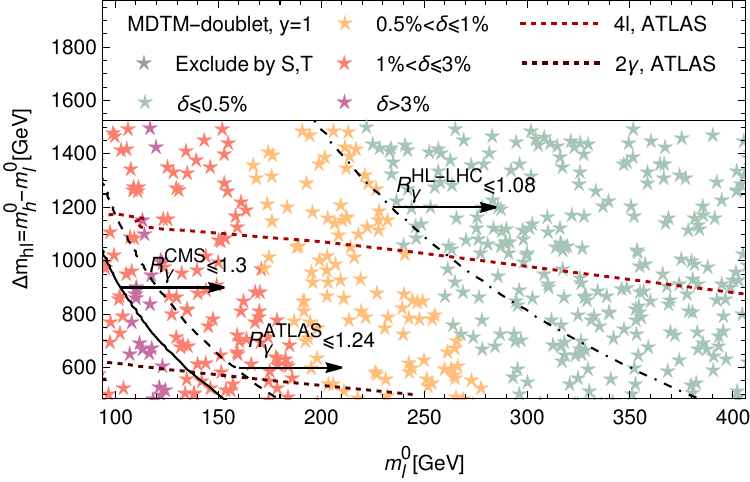}
    \\[1em]
    \includegraphics[width=0.65\textwidth]{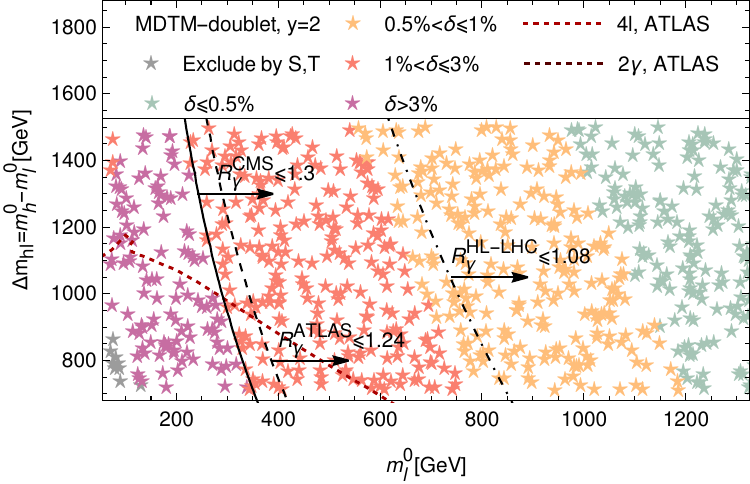}
    \caption{Parameter scan result for the MDTM with $y=1$(upper) and $y=2$(lower) in the doublet-dominant scenario. See caption of Fig.~\ref{fig:MSDMdec} for the definition of the curves. The black lines denote the upper limits of $R_\gamma$ from the CMS \cite{CMS:2021kom}, ATLAS \cite{ATLAS:2022tnm} as well as HL-LHC\cite{Cepeda:2019klc}, respectively.}
    \label{fig:MDTMdec}
\end{figure}

Finally, Fig.~\ref{fig:MDTMdec} shows results for the MDTM in the doublet-dominated scenario. In the scenario where $\chi^0_l$ decays into leptons, the LHC constraints are fairly strong, but there are extensive regions of parameter space with large $\Delta m_{hl} = m_h^0-m_l^0$ that are beyond the reach of the LHC while leading to significant deviations of $\sigma_{\rm ZH}$ up to several percent. For the scenario of $\chi^0_l$ decays into photons, the direct LHC search bounds are even more limited. Besides measurements of $\sigma_{\rm ZH}$ at future $e^+e^-$ colliders, the region with large $\Delta m_{hl}$ can also be probed, to a certain extent, with precision measurements of the Higgs decay rate to photons ($R_\gamma$) at the HL-LHC, as discussed in section~\ref{sec:MajDTModel}.
The situation is similar for the triplet-dominated scenario.

\begin{table}[h!]
    \centering
  \setlength\extrarowheight{6pt}
   \resizebox{\linewidth}{!}{ 
  \begin{tabular}{m{12mm} m{35mm} m{40mm} m{35mm} m{15mm} m{20mm}}
    \hline\hline
Model & Scenario  &  Main LHC prod.\ channel & (HL-)LHC constraint & $\delta\sigma^{y\leq 2}_{\rm ZH, max}$ & oblique para.  \\[0.5em] \hline 
DSDM & large $\Delta m_{Dl}$  & $pp\to\chi_D^\pm\chi_h^0$ & $pp\to 4q$, \cite{ATLAS:2021yqv} & $<3\%$ & relevant \\
 &  & & $pp\to 1lbb$, \cite{ATLAS:2018diz} & &  \\[-1em]
&  \multicolumn{5}{c}{\dotfill} \\[-0.5em]
 & small $\Delta m_{Dl}$  & $pp\to\chi_D^\pm\chi_D^\mp$ & $pp\to 2l_{\rm soft}$, \cite{Zhou:2022jgj} & $<1\%$ & relevant \\[0.5em]
\hline
MSDM   & $\chi_h^0$ doublet-dominant & $pp\to\chi_{h,D}^0\chi_D^\pm,\chi_D^\pm\chi_D^\mp$ & $pp\to 4q$, \cite{ATLAS:2021yqv} & $<6\%$ & not relevant \\
 &  & & $pp\to 1lbb$, \cite{ATLAS:2018diz} & &  \\[-1em]
&  \multicolumn{5}{c}{\dotfill} \\[-0.5em]
& $\chi_h^0$ singlet-dominant & $pp\to\chi_D^0\chi_D^\pm$  & $pp\to  3l_{\rm soft}$,\cite{ATLAS:2021moa} & $<1\%$ & not relevant \\
& & & $pp\to \leq2l_{\rm soft}$,\cite{Cardona:2021ebw} & & \\
\hline
DDTM1 & $\chi_l^\pm$ doublet-dominant  & $pp\to\chi_l^\pm\chi_l^\mp$ & $pp\to 2l_{\rm soft}$,\cite{Zhou:2022jgj} & $<3\%$ & relevant \\[-1em]
&  \multicolumn{5}{c}{\dotfill} \\[-0.5em]
 & $\chi_l^\pm$ triplet-dominant  & $pp\to\chi_l^\pm\chi_l^\mp$ & $pp\to 2l_{\rm soft}$, \cite{Zhou:2022jgj} & $<3\%$ & relevant \\
\hline
DDTM0 & $\chi_h^0$ triplet-dominant & $pp\to\chi_h^0\chi_{h,T}^\pm,\chi_{h,T}^\pm\chi_{h,T}^\mp$ & $pp\to 4q$, \cite{ATLAS:2021yqv} & $<1\%$ & relevant \\
 &  & & $pp\to 1lbb$, \cite{ATLAS:2018diz} & & \\[-1em]
&  \multicolumn{5}{c}{\dotfill} \\[-0.5em]
 & \multicolumn{2}{l} {$\chi_h^0$ doublet-dominant is forbidden} & & & \\
\hline
MDTM  & $\chi_h^0$ doublet-dominant & $pp\to\chi_{h,D}^0\chi_h^\pm,\chi_h^\pm\chi_h^\mp$  & $pp\to 4q$, \cite{ATLAS:2021yqv} & $<4\%$ & relevant \\
 &  & & $pp\to 1lbb$, \cite{ATLAS:2018diz} & &  \\[-1em]
&  \multicolumn{5}{c}{\dotfill} \\[-0.5em]
  & $\chi_h^0$ triplet-dominant & $pp\to\chi_h^0\chi_h^\pm,\chi_h^\pm\chi_h^\mp $ & $pp\to 4q$, \cite{ATLAS:2021yqv} & $<3\%$ & relevant \\
 &  & & $pp\to 1lbb$, \cite{ATLAS:2018diz} & &  \\
\hline \hline
    \end{tabular}}
    \caption{Summary table for the scenarios with stable lightest fermions considered in this work. $\delta\sigma^{y\leq 2}_{\rm ZH, max}$ denotes the size of the possible deviations of the $e^+e^- \to ZH$ cross-section within current constraints from electroweak precision and LHC data.}
    \label{tab:summary}
\end{table}

\section{Conclusions}
\label{sec:concl}
TeV-scale new particles could cause percent-level deviations in effective Higgs couplings, offering indirect evidence of new physics, which is the primary task of future Higgs factories, such as ILC, CEPC and FCC-ee. On the other hand, direct search for these particles can offer complementary information. In this study, we investigate the discovery potential of dark sector fermions through Higgs precision studies and collider searches, within simplified UV complete models. The models extend the SM by two weak SU(2) Majorana or Dirac fermionic multiplets, wherein the lightest charged neutral fermion can serve as a dark matter candidate (``minimal scenario'') or decay ( ``non-minimal scenarios''). Both of them are considered in this work. 

Specifically, we have considered five different weak gauge quantum number assignments of the new fermions: an SU(2) singlet plus an SU(2) doublet, where the singlet could be Majorana or Dirac; and an SU(2) doublet plus an SU(2) triplet, where the triplet could be Majorana or Dirac. For the Dirac triplet, we consider two options for its hypercharge, 0 and $-1$. All these models permit a Yukawa coupling between the two new fermion fields and the Higgs boson, which plays and important role for the Higgs phenomenology.

The cross-section for $e^+e^- \to ZH$ is modified through loop corrections involving the new fermions. There deviation would be observable at future Higgs factories if they reach the level of ${\cal O}(1\%)$. We find deviations of at least 0.5\% for large regions of parameter with new fermion masses of several hundred GeV, and in some cases extending beyond 1~TeV, and ${\cal O}(1)$ Yukawa couplings. However, large fractions of these parameter regions can also be probed through obliqure parameter constraints from LEP electroweak precision data, Higgs data from LHC (in particular the $H\to\gamma\gamma$ decay), as well as direct searches for the new fermions at the LHC. The latter leads to signatures that are very similar to gaugino production and decay in the MSSM (for the minimal scenario) or extended SUSY models with R-parity violation or light gravitinos (for the non-minimal scenarios). Therefore, existing limits from direct searches at the LHC and projections for the HL-LHC can be obtained by recasting gaugino search studies in the literature.

We have studied the complementarity of future precision measurements of the $e^+e^- \to ZH$ cross-section ($\sigma_{\rm ZH}$) and constraints from LEP and LHC for the five models listed above. Constraints from oblique parameters are mostly irrelevant for the Majonara models, whereas for the Dirac models they restrict the viable parameter space to relatively narrow slices where $\sigma_{\rm ZH}$ is modified by more than 0.5\% (but rarely more than $1\%$). The LHC phenomenology depends on whether one consider the minimal or non-minimal scenarios. 

Let us first summarize the results in the minimal scenario, where the lightest new fermion is stable and escapes from the detector as missing energy. For moderate Yukawa couplings ($y \lesssim 1$) and mass differences $\gtrsim 200$~GeV between the new fermions, direct (HL-)LHC searches, combined with $H \to \gamma\gamma$ constraints, can exclude major portions of the parameter space that is viable for observable $\sigma_{\rm ZH}$ effects in all models. For larger values of the Yukawa couplings, $y>1$, $\sigma_{\rm ZH}$ precision measurements can be sensitive to ${\cal O}$(TeV) masses of the new fermions, which is beyond the reach of the LHC, in particular for the Majorana models. For small mass differences, bounds from direct LHC searches becomes much weaker, whereas $\sigma_{\rm ZH}$ can deliver interesting information about this region. 

In the non-minimal case, we consider decays of the lightest new fermions into final states with leptons and into final states with photons. For large mass differences of the new fermions, the direct LHC bounds in the leptonic decay scenario are stronger than in the minimal scenario, whereas they are weaker in the photonic decay scenario. On the other hand, for small differences the LHC bounds in both the leptonic and photonic decay scenario are significantly stronger than for the minimal scenario, and they can exclude most of the viable parameter for $\sigma_{\rm ZH}$ measurements in that region. This can be easily understood because small mass differences among the new fermions implies that their visible decay products are soft and cannot be used for triggering at the LHC. Therefore, in the minimal scenario a hard jet is typically required for triggering, which reduces the signal rate, whereas in the non-minimal scenarios the additional leptons or photons from the decay of the lightest new fermion can used instead.

Overall we conclude that both the (HL-)LHC and future $e^+e^-$ Higgs factories are powerful tools to study models with new electroweak fermions, and they complement each other. Depending on the specific model and size of the new Yukawa coupling, the LHC sensitivity can dominate in some cases, whereas the Higgs factories have superior prospects in other cases. Finally, it should be that improved determinations of electroweak precision quantities at future $e^+e^-$ collider, and thus the oblique parameters, would provide an alternative way of probing these dark sector fermions. In fact, the pattern of deviations in different observables in the Higgs and electroweak sectors could help to discriminate the type of new physics underlying these effects. We leave a study of this question for future work.

\section*{Acknowledgments}

This work has been supported in part by the National Science Foundation under grant no.~PHY-2112829.

\pagebreak[0]

%%%%%%%%%%%%%%%%%%%%%%%%%%%%%%%%%%%%%%%%%%%%%%%%%%%%%%%%%%%%%%

%\bibliographystyle{utcaps_mod}
\bibliography{dmferm}

\end{document}